\documentstyle[nato]{crckapb} 

\begin{opening}
\title{SPECTRA OF LARGE RANDOM MATRICES:\protect\\
       A METHOD OF STUDY}

\author{E. KANZIEPER}
\institute{The Abdus Salam International Centre for Theoretical Physics\\
           P.O.B. 586, 34100 Trieste, Italy}
\author{V. FREILIKHER}
\institute{The Jack and Pearl Resnick Institute for Advanced Technology\\
           Department of Physics, Bar--Ilan University\\
           52900 Ramat--Gan, Israel} 
\end{opening}

\begin{document}

\begin{abstract}
A formalism for study of spectral correlations in non--Gaussian,
unitary invariant ensembles of large random matrices with strong level
confinement is reviewed. It is based on the Shohat method in the theory of
orthogonal polynomials. The approach presented is equally suitable for
description of both local and global spectral characteristics, thereby
providing an overall look at the phenomenon of spectral universality in
Random Matrix Theory.
\end{abstract}

\section{Introduction: Motivation and Basic Results}

\subsection{Ubiquity of Invariant Random Matrix Models}

Random matrices \cite{M-1991,GMGW-1998} have been introduced in a physical
context since the pioneering works by Wigner \cite{W-1951} and Dyson \cite
{D-1962,D-1962a}. Initially proposed as an effective phenomenological model
for description of the higher excitations in nuclei \cite{BFFMPW-1981} they
found numerous applications in very diverse fields of physics such as two
dimensional quantum gravity \cite{FGZJ-1995}, quantum chromodynamics \cite
{VZ-1993}, quantum chaos \cite{G-1990}, and mesoscopic physics \cite
{ALW-1991,B-1997}. One can state that from the standpoint of the
mathematical formalism all these fields are pooled by the Random Matrix
Theory (RMT). Such an ubiquity of random matrices owes its origin to the
exclusive role played by symmetry. The most amazing evidence to this fact
comes from {\it invariant random matrix models }which will be the focus of
this review. The main feature of these models lies in that they discard
(irrelevant) microscopic details of the physical system in question, but
they do properly take into account its underlying fundamental symmetries. In
accordance with the very idea of the construction of invariant matrix
models, they do {\it not} relate to any dynamical properties of the physical
object under study: general symmetry requirements alone lead to appearance
of knowledge about the system. As far as the symmetry constraints follow
from the first principles, the Random Matrix Theory turns out to be a
general and powerful field--theoretical approach leading to a unified
mathematical description of the quite different physical problems mentioned
above.

A great variety of invariant random matrix ensembles can be assigned to
three irreducible symmetry classes \cite{D-1962a}. To specify them, we
consider the typical line of arguing used in applications of the Random
Matrix Theory to the disordered quantum mechanical systems, where it was
first invented by Gor'kov and Eliashberg \cite{GE-1965}. Since in this
situation the microscopic Hamiltonian ${\cal H}$ is rather intricate, the
integration of exact equations is impossible. It is therefore useful to
appeal to statistical description by conjecturing that the operator ${\cal H}
$ can be modelled by an $N\times N$ random matrix ${\bf H}$ whose
eigenvalues and eigenvectors reproduce statistically the eigenlevels and
eigenfunctions of the real microscopic Hamiltonian in the thermodynamic
limit which corresponds to the matrix dimension $N$ going to infinity. With
this conjecture accepted, an ensemble of large random matrices ${\bf H}$,
characterized by the joint distribution function $P\left[ {\bf H}\right] $
of the matrix elements ${\bf H}_{ij}$ of the corresponding Hamiltonian $%
{\cal H}$, becomes the main object of study. Once the primary role of
symmetry is postulated in the RMT--approach, the matrix ${\bf H}$ must
adequately reflect the symmetry properties of the physical system under
study. The matrix ${\bf H}$ is chosen to be real symmetric if the underlying
physical system possesses time--reversal and rotational invariance. Systems
with broken time--reversal symmetry are characterized by a Hermitean matrix $%
{\bf H}$, while systems with conserved time--reversal symmetry but with
broken rotational invariance are described by a self--dual Hermitean matrix.
These three symmetry classes referred to as the orthogonal, unitary and
symplectic symmetry classes, respectively, can be characterized by a
symmetry parameter $\beta $ equals the number of independent elements in the
off--diagonal entries of the matrix ${\bf H}$. The parameter $\beta =1$ for
a real symmetric matrix (orthogonal symmetry), $\beta =2$ for a Hermitean
matrix (unitary symmetry), and $\beta =4$ for a self--dual Hermitean matrix
(symplectic symmetry).

So far we did not yet specify the form of the joint distribution function $%
P\left[ {\bf H}\right] $ of the matrix elements ${\bf H}_{ij}$. By
definition, the invariant random matrix ensembles are characterized by 
\begin{equation}
P\left[ {\bf H}\right] =\frac 1{{\ }{\cal Z}_N}\exp \left\{ -\beta {\rm 
\mathop{\rm Tr}
}V\left[ {\bf H}\right] \right\} .  \label{mar.00}
\end{equation}
Here the function $V\left[ {\bf H}\right] $ should ensure the existence of
the partition function ${\cal Z}_N$, which is defined by the normalization
condition $\int P\left[ {\bf H}\right] d\left[ {\bf H}\right] =1$ with the
elementary volume $d\left[ {\bf H}\right] $ depending on the symmetry of the
matrix ${\bf H}$. For $\beta =1$ the volume element $d\left[ {\bf H}\right] =\prod_{i\leq j}d{\bf H}_{ij}$, for $\beta =2$ the volume element $d\left[ {\bf H}\right] =\prod_{i\leq j}d%
\mathop{\rm Re}
{\bf H}_{ij}\prod_{i<j}d{\rm 
\mathop{\rm Im}
}{\bf H}_{ij}$, while for $\beta =4$ it equals
$d\left[ {\bf H}\right] =\prod_{i\leq j}d{\bf H}_{ij}^{\left( 0\right)
}\prod_{\sigma =1}^3\prod_{i<j}d{\bf H}_{ij}^{\left( \sigma \right) }$. %
 The presence of the trace in Eq. (\ref{mar.00}) leads to the invariance of the probability density 
$P\left[ {\bf H}\right] d\left[ {\bf H}\right] $ under the similarity
transformation ${\bf H}\rightarrow {\cal R}_\beta ^{-1}{\bf H}{\cal R}_\beta 
$ with ${\cal R}_\beta $ being an orthogonal, unitary or symplectic $N\times
N$ matrix for $\beta =1,2$ or $4$, respectively. In turns, the invariance
built in the probability density $P\left[ {\bf H}\right] d\left[ {\bf H}%
\right] $ implies that {\it there is no preferential basis in the space of
matrix elements}. From the physical point of view, this means that invariant
matrix models given by Eq. (\ref{mar.00}) are applicable to particular
regimes of a physical system where (i) all the normalized linear
combinations of the eigenstates have similar properties, and where (ii) the
dimensionality is irrelevant. In disordered systems this is just a metallic
state where the typical electron states are extended and hence structureless.

Notice that up to this point we have no constraints allowing us to uniquely
choose the function $V\left[ {\bf H}\right] $ (referred to as ``confinement
potential''). However, if we impose the additional requirement that the
entries of the random matrix ${\bf H}$ be statistically independent of each
other, we immediately arrive at the Gaussian Orthogonal, Unitary and
Symplectic Ensembles (GOE, GUE, GSE) which are characterized by the
quadratic confinement potential $V\left[ {\bf H}\right] \propto {\bf H}^2$.
This particular form of confinement potential leads to significant
mathematical simplifications which allowed the complete treatment of these
three ensembles many years ago \cite{P-1965}.

It is remarkable that (even for non--Gaussian distributions of $P\left[ {\bf %
H}\right] $) the invariant random matrix ensembles possess a great degree of
mathematical tractability, and, what is more important, they have a high
physical relevance, being much more than just a mathematical construction.
In this respect we mention that in the physics of disordered systems the
applicability of Gaussian invariant random matrix ensembles to description
of weakly disordered systems has been proven by Efetov \cite{E-1983} by
using the nonlinear $\sigma $--model. In that study \cite{E-1983} the
statistical properties of energy levels for metallic particles with volume
imperfections were considered by solving the Schr\"{o}dinger equation with
nonperturbative averaging over the random potential configurations within
the framework of the supersymmetry method. Random Matrix Theory appears
there as a zero--dimensional version of a more general microscopic nonlinear 
$\sigma $--model, thereby proving the validity of the basic principles used
for an RMT phenomenological description of energy levels of noninteracting
electrons confined in a restricted volume. This connection takes place at
times much larger than the ergodic time needed for diffusive particle to
completely and homogeneously fill the available volume of the sample
provided it is in the metallic regime characterized by a dimensionless
Thouless conductance $g\gg 1$.

Let us point out that the choice of quadratic confinement potential $V\left[ 
{\bf H}\right] $ can hardly be justified. Indeed, it was understood from the
very beginning \cite{P-1965} that the requirement of statistical
independence of the matrix elements ${\bf H}_{ij}$ is not motivated by the
first principles and, therefore, the important problem of elucidating the
influence of a particular form of confinement potential on the predictions
of the Random Matrix Theory developed for Gaussian Invariant Ensembles had
been posed already in the sixties. Despite this fact, considerable progress
in study of spectral properties of {\it non--Gaussian} random matrix
ensembles was achieved almost thirty years later when RMT experienced a great
renaissance due to new ideas in the physics of disordered/chaotic systems
which had led the birth of mesoscopic physics, as well as due to a
penetration of Random Matrix Theory to quantum chromodynamics (QCD). In the
latter field, the Random Matrix Theory turned out to be a useful tool for
understanding the spectral properties of low--lying eigenvalues of the Dirac
operator. The idea of introducing the RMT--approach in QCD is very similar
to that in the physics of disordered systems and is based on the conjecture 
\cite{VZ-1993,BBMSVW-1998} that the spectral density of the Dirac operator
very close to the spectrum origin should depend only on the symmetries in
question. One startling consequence of this conjecture is that the spectral
density of the Dirac operator near the origin need not be computed within
the framework of the gauge theories at all, but it can be extracted from
much simpler random matrix ensembles reflecting the symmetry of the problem.
The matrix models appearing in the context of QCD are manifestly
non--Gaussian possessing an additional chiral structure\footnote{%
Throughout the paper we consider non--chiral non--Gaussian random matrix
ensembles. It can be shown that an arbitrary chiral matrix
model can be reduced to an auxiliary one without chirality; see, for instance,
Ref. \cite{ADMN-1997}.} \cite{QCD-1998}. Another
motivation for studying the spectral properties of non--Gaussian random
matrix ensembles comes from the theories of $2D$ quantum gravity \cite
{FGZJ-1995}.

\subsection{Non--Gaussian Random Matrix Ensembles and Phenomenon of Spectral
Universality}

The examples above clearly demonstrate an important role the non--Gaussian
random matrix ensembles play in different physics theories, and serve as a
compelling evidence of the necessity to have a powerful method for study of
their spectral properties which could be equally applicable to rather
different probability measures $P\left[ {\bf H}\right] $. During (mostly)
the recent decade a number of methods were developed in order to treat
non--Gaussian random matrix ensembles. All of them can schematically be
related to two groups.

\subsubsection{Global Universality}

The first group includes different approximate methods useful to explore 
{\it global spectral characteristics} which manifest themselves on the scale
of $n\gg 1$ eigenlevels. Among these methods there are (i) the mean--field
approximation proposed by Dyson \cite{D-1962,D-1972} which allows one to
compute the density of levels in random matrix ensemble; (ii) the
Schwinger--Dyson loop equations' technique \cite{AJM-1990,AA-1996,I-1997}
that had led to discovery of the phenomenon of the {\it global spectral
universality};{\it \ }(iii) the method of functional derivative of Beenakker 
\cite{B-1993,B-1994}, and (iv) the diagrammatic approach of Br\'{e}zin and
Zee \cite{BZ-1994} whose development enabled their authors to study the
phenomenon of global universality \cite{AJM-1990} in more detail as well as
to generalize it in the context of mesoscopic physics.

It was found that contrary to the one--point spectral characteristics (such
as one--point Green's function or level density) which essentially depend on
the measure $P\left[ {\bf H}\right] $, i.e. on the explicit form of the
confinement potential, the {\it functional form} of (connected) two--point
correlators becomes insensitive to the details of confinement potential upon
smoothing over the scale which is much larger than the mean level spacing $%
\Delta _N$ but much smaller than the scale of the entire spectrum support.
This is the essence of the phenomenon of global spectral universality. For
example, the smoothed connected density--density correlator computed for the
matrix model Eq. (\ref{mar.00}) is given by the universal function 
\begin{equation}
\rho _c^{\left( N\right) }\left( \lambda ,\lambda ^{\prime }\right) =-\frac 1%
{\pi ^2\beta \left( \lambda -\lambda ^{\prime }\right) ^2}\frac{{\cal D}%
_N^2-\lambda \lambda ^{\prime }}{\left( {\cal D}_N^2-\lambda ^2\right)
^{1/2}\left( {\cal D}_N^2-\lambda ^{\prime 2}\right) ^{1/2}},  \label{gu}
\end{equation}
where $\lambda \neq \lambda ^{\prime }$. This form of $\rho _c^{\left(
N\right) }$ is valid for random matrices whose spectrum is supported on a
single, symmetric interval $\left( -{\cal D}_N,+{\cal D}_N\right) $. Here,
the universality implies that all the information about the particular form
of $V\left[ {\bf H}\right] $ is encoded into $\rho _c^{\left( N\right) }$
only through the end point ${\cal D}_N$ of the spectrum support.

The methods mentioned above, being applicable to study of the
spectral correlations in the long--range regime for all three symmetry
classes, leave aside the fine structure of eigenvalue correlations
manifested on the scale of the mean eigenvalue spacing. In this sense, these
approaches are less informative as compared to the method of orthogonal
polynomials \cite{M-1991} which, along with the supersymmetry approach \cite
{HW-1995,VWZ-1985}, enters the second group.

\subsubsection{Local Universality}

At present, the orthogonal polynomial technique, originally developed by
Gaudin and Mehta \cite{GM-1960}, seems to be the most powerful one
furnishing us the possibility to probe both the global spectral
characteristics (which are of importance in computing the integral spectral
properties) and the local spectral correlations (which describe a dynamics
of underlying physical system) for arbitrary symmetry classes.

(i) The early attempts \cite{FK-1964,L-1964,B-1965} to go beyond the
Gaussian distribution of $P\left[ {\bf H}\right] $ were concentrated on
unitary invariant, $%
\mathop{\rm U}
\left( N\right) $, matrix ensembles ($\beta =2$) associated with classical
orthogonal polynomials. It was found in Refs. \cite
{FK-1964,L-1964,B-1965,NW-1991} that (in the thermodynamic limit $%
N\rightarrow \infty $) the scalar two--point kernel (in terms of which all $n
$--point correlation functions are expressible, see Sec. 2.2) computed in
the bulk of the spectrum, i.e. far from the end points of the eigenvalue
support, follows the {\it sine law} 
\begin{equation}
K_{\rm{bulk}}\left( s,s^{\prime }\right) =\frac{\sin \left[ \pi \left(
s-s^{\prime }\right) \right] }{\pi \left( s-s^{\prime }\right) },  \label{sk}
\end{equation}
that inevitably leads to the famous Wigner--Dyson level statistics \cite
{BTW-1992} inherent in nuclear physics, weakly disordered and chaotic
systems. This form of the scalar kernel corresponds to the {\it bulk scaling
limit}, when the initial spectral variables $\lambda $, $\lambda ^{\prime }$
are measured in the units of the mean eigenvalue spacing $\Delta _N$, so
that the scaled variable $s=\lambda /\Delta _N$. The same form of the scalar
kernel Eq. (\ref{sk}) obtained for different ensembles associated with
classical orthogonal polynomials was the first evidence to the phenomenon
currently known as the {\it local spectral universality.} Later, it was
shown in Ref. \cite{NW-1991} that spectral properties of orthogonal $%
\mathop{\rm O}
\left( N\right) $ and symplectic $%
\mathop{\rm Sp}
\left( N\right) $ ($\beta =1$ and $\beta =4$) invariant ensembles associated
with the weights of classical orthogonal polynomials\footnote{%
In this situation the spectral correlations are expressible through the $%
2\times 2$ matrix kernel, that can be computed by using so--called skew
orthogonal polynomials \cite{BN-1991}. In fact, as was recently shown in
Ref. \cite{TW-1998}, the ensembles with $\beta =1$ and $\beta =4$ can be
also treated without appealing to the skew polynomials.} follow the
predictions of GOE and GSE, respectively.

A further significant progress in the field came with the works \cite
{MM-1991,BZ-1993} whose authors considered spectral properties of $%
\mathop{\rm U}
\left( N\right) $ invariant random matrix ensembles associated with strong
symmetric confinement potentials of the form $V\left[ {\bf H}\right] ={\bf H}%
^2+\gamma {\bf H}^4$ $\left( \gamma >0\right) $ and $V\left[ {\bf H}\right]
=\sum_{k=1}^pa_k{\bf H}^{2k}$ $\left( a_p>0\right) $, respectively. Both
works, based on different conjectures about the functional form of
asymptotics of polynomials orthogonal with respect to a non--Gaussian
measure, restricted their attention to the spectrum bulk, where the
two--point kernel was shown, once again, to follow the sine law, Eq. (\ref
{sk}). A rigorous treatment of a richer class of $%
\mathop{\rm U}
\left( N\right) $ invariant random matrix ensembles related to the Freud and
Erd\"{o}s--type orthogonal polynomials was given in Refs. \cite
{FKY-1996,FKY-1996a}. It was proven there that the universal sine law for
the two--point kernel holds for a wide class of monotonic (not necessarily
of polynomial form) non--singular confinement potentials $V\left( \lambda
\right) $ which increase at least as $\left| \lambda \right| $ at infinity,
and can grow as or even faster than any polynomial at infinity. Confinement
potentials satisfying these properties are referred to as {\it strong
confinement potentials}. This definition takes its origin in the limits \cite
{FKY-1996} of spectral universality and is non--accidently connected to the
problem of determinate and indeterminate moments \cite{ST-1963}. Also, it
was demonstrated in Ref. \cite{FKY-1996a} that an intimate connection exists
between the structure of Szeg\"{o} functional \cite{S-1921} entering the
strong pointwise asymptotics of orthogonal polynomials and the mean--field
equation by Dyson for mean level density that has been derived in Ref. \cite
{FKY-1996a} within the framework of orthogonal polynomial technique.

(ii) All the random matrix ensembles treated in Refs. \cite
{FK-1964,L-1964,B-1965,NW-1991,MM-1991,BZ-1993,FKY-1996,FKY-1996a} were
characterized by strong confinement potential with no singularities.
However, (logarithmic) singularities do appear when one considers chiral
matrix ensembles, arising in the context of QCD \cite{VZ-1993}, in the
theory of mesoscopic electron transport \cite{SN-1993} and in description of
electron level statistics in normalconducting--superconducting hybrid
structures \cite{AZ-1997}. An example of a rather general (though
non--chiral) random matrix ensemble possessing a log--singular level
confinement is given by the distribution 
\begin{equation}
P\left[ {\bf H}\right] =\frac 1{{\ }{\cal Z}_N}\left| \det {\bf H}\right|
^{\alpha \beta }\exp \left\{ -\beta {\rm 
\mathop{\rm Tr}
}V\left[ {\bf H}\right] \right\} ,  \label{1s}
\end{equation}
where the function $V\left[ {\bf H}\right] $ is a well behaved function
which has not singular points. Ensemble Eq. (\ref{1s}), being a natural
generalization of the matrix ensemble proposed by Bronk \cite{B-1965}, was
first considered in the situations where associated orthogonal polynomials
were classical \cite{B-1965,NS-1993,F-1993,NF-1995}. For quadratic
confinement potential and $\beta =2$ one obtains that in the vicinity of the
singularity, $\lambda =0$, the scalar kernel satisfies the {\it Bessel law} 
\begin{equation}
K_{\rm{orig}}\left( s,s^{\prime }\right) =\frac \pi 2\left( ss^{\prime
}\right) ^{1/2}\frac{J_{\alpha +1/2}\left( \pi s\right) J_{\alpha
-1/2}\left( \pi s^{\prime }\right) -J_{\alpha -1/2}\left( \pi s\right)
J_{\alpha +1/2}\left( \pi s^{\prime }\right) }{s-s^{\prime }},  \label{bk}
\end{equation}
where $s$ and $s^{\prime }$ are scaled by the level spacing $\Delta _N\left(
0\right) $ near the spectrum origin, $s=\lambda /\Delta _N\left( 0\right) $,
and $\alpha >-1/2$. This scaling procedure is referred to as the {\it origin
scaling limit.} Extensions to two other symmetry classes, as well as to the
chiral matrix ensembles, can be found in Refs. \cite
{NS-1993,F-1993,NF-1995,TW-1994}. An important breakthrough in understanding
the universal character of this kernel was given in Refs. \cite
{N-1996,ADMN-1997}. These authors, guided by QCD applications, have shown
that the Bessel kernel Eq. (\ref{bk}) is again universal, being independent
of the details of strong confinement potential $V\left[ {\bf H}\right] $. An
alternative proof of universality that holds more generally was presented in
Ref. \cite{KF-1998}.

(iii) The third type of universal correlations takes place near the soft
edge of the spectrum support, which is of special interest in the models of
two--dimensional quantum gravity. The first study of the level density near
the end point of the spectrum support is due to Wigner \cite{W-1962}. More
comprehensive description of the tails of the density of states was done in
Ref. \cite{BB-1991}. It was shown there that at $\beta =2$ a universal
crossover occurs from a nonzero density of states to a vanishing one that is
independent of the confining potential in the {\it soft--edge scaling limit. 
}Later it was demonstrated \cite{KF-1997} that eigenvalue {\it correlations}
in the $%
\mathop{\rm U}
\left( N\right) $ invariant matrix ensembles with quartic and sextic
confinement potentials are determined by the scalar kernel obeying the {\it %
Airy law }\cite{TW-1994a} 
\begin{equation}
K_{\rm{soft}}\left( s,s^{\prime }\right) =\frac{%
\mathop{\rm Ai}
\left( s\right) 
\mathop{\rm Ai}
^{\prime }\left( s^{\prime }\right) -%
\mathop{\rm Ai}
\left( s^{\prime }\right) 
\mathop{\rm Ai}
^{\prime }\left( s\right) }{s-s^{\prime }},  \label{ak}
\end{equation}
suggesting that the Airy kernel should be universal as well. (Here the
rescaling $s\propto N^{2/3}\left( \lambda /{\cal D}_N-1\right) $ determines
the soft--edge scaling limit). This conjecture has been proven in Ref. \cite
{KF-1997a}, where it was also shown that the Airy correlations, being
universal for a class of matrix models with monotonic confinement potential
or with that having light local extrema, are indeed a particular case of
more general universal multicritical correlations \cite{KF-1997a}.

\subsubsection{Universality in a Broader Context}

Are the universal scalar kernels in unitary invariant random matrix models
with strong level confinement exhausted by the universal sine, Bessel and
Airy laws given by Eqs. (\ref{sk}), (\ref{bk}) and (\ref{ak}) above? The
present state of the art leads us to the negative answer. Indeed, by adding
a singular component to the strong level confinement we may either
accumulate a finite number of eigenvalues near the singular point $\lambda _{%
\rm{sing}}$ or repel them from it. Such a rearrangement of eigenlevels
will lead to emergence of a new scalar two--point kernel in the vicinity of
the singular point of the spectrum. (In particular case of the logarithmic
singularity this kernel will follow the Bessel law Eq. (\ref{bk})).
Generically, it is natural to expect that the functional form of the kernel
near $\lambda _{\rm{sing}}$ will be sensitive to the particular type of
the singular deformation. However, the new two--point kernel will be
insensitive to the details of the background component of the confinement
potential, as it takes place in the case of the Bessel kernel. In this
sense, one can still say about a phenomenon of universality. One of the
latest evidences to this fact can be found in Ref. \cite{DN-1998} where
deformed ensembles of large random matrices associated with massive Dirac
operators were considered.

By the same token, the universal global spectral correlator expressed by Eq.
(\ref{gu}) is not the only one arising in Random Matrix Theory. As it was
already stressed, the universal function Eq. (\ref{gu}) is inherent in
matrix models with spectra possessing a single connected support.
Correspondingly, ensembles of large random matrices with eigenvalue
densities having more than one--cut support will give rise to the novel
global spectral correlators whose universality classes (for a given symmetry
parameter $\beta $) are characterized entirely by the number of cuts in the
support of spectral density. This was explicitly demonstrated in the recent
studies \cite{AA-1996,A-1996} by means of the loop equation technique.

\subsection{The Aim}

At this point, it is appropriate to notice that all the mentioned above
(universal) results for both global and local eigenvalue correlations have
been obtained by using different methods, each of them was only suitable for
a particular problem under consideration. Any deformations of $P\left[ {\bf H%
}\right] $ (which will preserve its invariance) would cause principal
difficulties in elucidating the influence of these deformations on spectral
properties of corresponding random matrix ensembles. The goal of this paper
is to represent a general method (recently developed in a series of
publications \cite{KF-1997a,KF-1998,KF-1998a}), which is equally suitable
for study of both local and global eigenvalue correlations, and easily leads
to generalizations. The approach we introduce is based on a simple and
elegant idea \cite{S-1939} by J. Shohat (which goes back to 1930), providing
a detailed description of the spectral properties of non--Gaussian $%
\mathop{\rm U}
\left( N\right) $ invariant random matrix ensembles through the analysis of the three--term
recurrence equation for associated orthogonal polynomials. We show that for
the most situations of interest, the knowledge of the large--$N$ behavior of
the coefficients in the recurrence equation is sufficient to directly
reconstruct the local eigenvalue correlations of arbitrary order, as well as
to explore the global spectral statistics. In the case of a non--singular,
well behaved confinement potential, the knowledge of such a large--$N$
behavior of the recurrence coefficients is equivalent, in fact, to a
knowledge of the Dyson density of states for the corresponding random matrix
ensemble. The latter assertion leads to a rather unexpected conclusion: Once
the Dyson density of states (which is a rather crude one--point spectral
characteristics) is available, the scalar kernel (and hence the $n$--point
spectral correlators) can immediately be recovered through the solution of a
certain second--order differential equation (See Sec. 4.3).

We believe that this method offers not only new computational
potentialities, but also provides a different, overall look at the problem
of eigenvalue correlations in unitary invariant random matrix ensembles in
arbitrary spectrum range and in arbitrary scaling limits. It seems that
together with the very recent works \cite{SV-1998,W-1998} establishing a
precise connection of the scalar kernel for random matrix ensembles with $%
\mathop{\rm U}
\left( N\right) $ symmetry with the $2\times 2$ matrix kernels in ensembles
with $%
\mathop{\rm O}
\left( N\right) $ and $%
\mathop{\rm Sp}
\left( N\right) $ symmetries, the formalism to be reviewed below gives a
rather complete solution of the problem of eigenvalue correlations in
invariant matrix models with strong level confinement.

The review is organized as follows. Section 2 contains a brief description
of the Gaudin--Mehta calculational scheme, that introduces the orthogonal
polynomials as a tool for exact evaluation of $n$--point correlation
functions in $%
\mathop{\rm U}
\left( N\right) $ invariant random matrix ensembles. In Section 3 the
Shohat method in the theory of orthogonal polynomials is presented. Section
4 is devoted to a detailed study of spectral properties of large Hermitean
random matrices with a single connected eigenvalue support. In Section 5 we extend this analysis to random
matrices with eigenvalue gap. Section 6 contains conclusions. The most
lengthy calculations are collected in three Appendices.

\section{Elements of the Gaudin--Mehta formalism}

\subsection{Invariant Random Matrix Model in Eigenvalue Representation: Two
Interpretations}

The invariance of the distribution function $P\left[ {\bf H}\right] $
implies that different matrices with the same eigenvalues have the same
probability of occurring. To study spectral characteristics of an invariant
random matrix model it is convenient to integrate out ``auxiliary'' angular
variables in the construction $P\left[ {\bf H}\right] d\left[ {\bf H}\right] 
$ in order to get the matrix model in the eigenvalue representation. To
proceed with this, we have to pass from the integration over independent
elements ${\bf H}_{ij}$ of the matrix ${\bf H}$ to the integration over the
smaller space of its $N$ eigenvalues $\left\{ \lambda \right\} $,
calculating the corresponding Jacobian $J$.

Let us introduce the matrix ${\cal R}_\beta $ that diagonalizes the random
matrix ${\bf H}$, so that ${\bf H}={\cal R}_\beta ^{-1}\Lambda {\cal R}%
_\beta $ and $\Lambda =%
\mathop{\rm diag}
\left( \lambda _1,...,\lambda _N\right) $, and consider the infinitesimal
variation of ${\bf H}$, 
\begin{equation}
\delta {\bf H=}{\cal R}_\beta ^{-1}\left( {\cal R}_\beta \delta {\cal R}%
_\beta ^{-1}\Lambda +\delta \Lambda +\Lambda \delta {\cal R}_\beta {\cal R}%
_\beta ^{-1}\right) {\cal R}_\beta ={\cal R}_\beta ^{-1}\left( \delta
\Lambda -i\left[ \Lambda ,\delta s\right] \right) {\cal R}_\beta. 
\label{rm.02}
\end{equation}
Here we have denoted $\delta s=i\delta {\cal R}_\beta {\cal R}_\beta ^{-1}$.
The norm, $\left\| \delta {\bf H}\right\| ^2=%
\mathop{\rm Tr}
\left( \delta {\bf H}\right) ^2$, of the infinitesimal variation of ${\bf H}$
is 
\begin{equation}
\left\| \delta {\bf H}\right\| ^2=%
\mathop{\rm Tr}
\left( \delta \Lambda \right) ^2-2i%
\mathop{\rm Tr}
\left( \left[ \delta \Lambda ,\Lambda \right] \delta s\right) +2%
\mathop{\rm Tr}
\left( -\delta s\Lambda \delta s\Lambda +\left( \delta s\right) ^2\Lambda
^2\right) .  \label{rm.03}
\end{equation}
The second term in the last expression vanishes as the matrices $\delta
\Lambda $ and $\Lambda $ are diagonal. We then find 
\begin{equation}
\left\| \delta {\bf H}\right\| ^2=\sum_i\left( \delta \lambda _i\right)
^2+\sum_{i,j}\left( \lambda _i-\lambda _j\right) ^2\left| \delta
s_{ij}\right| ^2.  \label{rm.04}
\end{equation}
The independent variables are the variations of the eigenvalues $\delta
\lambda _i$ and $\delta s_{ij}$ for $\beta =1$, $%
\mathop{\rm Re}
\delta s_{ij}$, $%
\mathop{\rm Im}
\delta s_{ij}$ for $\beta =2$, or $\delta s_{ij}^{\left( \sigma \right) }$
with $\sigma =0,1,2,3$ for $\beta =4$, $i<j$. Then, from Eq. (\ref{rm.04})
we obtain the Jacobian $J$ of the transformation ${\bf H\mapsto }\left\{
\lambda _i,{\cal R}_\beta \right\} $ that is equal to $\sqrt{\det G}$, where 
$G$ is the metric tensor with $\det G=\prod_{i\neq j}\left| \lambda
_i-\lambda _j\right| ^\beta $. Hence 
\begin{equation}
J=\prod_{i<j}\left| \lambda _i-\lambda _j\right| ^\beta =\left| \Delta
\left( \lambda \right) \right| ^\beta  \label{rm.05}
\end{equation}
with $\Delta \left( \lambda \right) $ being the Vandermonde determinant. The
construction Eq. (\ref{rm.05}) is also known as the Jastrow factor.

Combining Eqs. (\ref{mar.00}) and (\ref{rm.05}) we arrive at the famous
expression for the joint probability density function of the eigenvalues $%
\left\{ \lambda \right\} $ of the matrix ${\bf H}$: 
\label{rm.06}
\begin{eqnarray}
\rho \left( \lambda _1,\ldots ,\lambda _N\right)  &=&{\cal Z}_N^{-1}\exp
\{-\beta \sum_{i=1}^NV\left( \lambda _i\right) \}\left| \Delta \left(
\lambda \right) \right| ^\beta   \label{rm.06a} \\
&=&{\cal Z}_N^{-1}\exp \{-\beta [\sum_iV\left( \lambda _i\right)
-\sum_{i<j}\ln \left| \lambda _i-\lambda _j\right| ]\}. 
\label{rm.06.b}
\end{eqnarray}
The probability distribution given by Eq. (\ref{rm.06.b}) has the form of a Gibbs
distribution for a classical one--dimensional system of $N$ ``particles'' at
``positions'' $\lambda _i$ confined by the external one--body potential $%
V\left( \lambda \right) $ and {\it interacting} with each other through the
pairwise logarithmic law originated from the Jacobian $J$ of the
transformation ${\bf H\mapsto }\left\{ \lambda _i,{\cal R}_\beta \right\} $.
The symmetry parameter $\beta $ plays the role of the equilibrium
temperature. Such an interpretation of Eq. (\ref{rm.06.b}) gives rise to
approximate mean--field methods in the Random Matrix Theory.

For  invariant matrix ensembles with unitary symmetry $(\beta =2)$, the probability
distribution in the form of Eq. (\ref{rm.06a}) can alternatively be related to a system of
fictitious {\it non--interacting fermions}, that can be described by
effective one--particle Schr\"{o}dinger equation \cite{KF-1997a}. This
equation is a cornerstone of the method under review. Although such a
simple, transparent interpretation cannot be ascribed to $\rho \left(
\lambda _1,...,\lambda _N\right) $ for two other symmetry classes, $\beta =1$
and $\beta =4$, a recently discovered deep connection \cite{SV-1998,W-1998}
between orthogonal, unitary and symplectic ensembles makes the unitary
invariant ensembles of random matrices to be the central and most important.

\subsection{Orthogonal Polynomials' Technique: $\beta =2$}

For $\beta =2$ the structure of Eq. (\ref{rm.06a}) enables us to exactly
represent all the statistical characteristics of the spectrum in the terms
of polynomials orthogonal with respect to a non--Gaussian measure $%
d\alpha$. To demonstrate this, it is useful to write down the
joint probability density function of the eigenvalues $\left\{ \lambda
\right\} $ in the form 
\label{rm.006}
\begin{eqnarray}
\rho \left( \lambda _1,\ldots ,\lambda _N\right)  &=&\Psi _0^2\left( \lambda
_1,\ldots ,\lambda _N\right) ,  \label{req.06a} \\
\Psi _0\left( \lambda _1,\ldots ,\lambda _N\right)  &=&{\cal Z}_N^{-1/2}\exp
\{-\sum_{i=1}^NV\left( \lambda _i\right) \}\Delta \left( \lambda \right) .
\label{req.06b}
\end{eqnarray}
One can see that $\Psi _0$ can be thought of as a wave function of $N$
fictitious non--interacting fermions. Namely, noting that 
\begin{equation}
\Delta \left( \lambda \right) =\prod_{i>j=1}^N\left( \lambda _i-\lambda
_j\right) =\det \left\| \lambda _i^{j-1}\right\| ,  \label{req.6&7}
\end{equation}
and taking the linear combinations of the columns of the initial matrix with
entries $\lambda _i^{j-1}$, one can reduce this matrix to the matrix whose
entries are arbitrary polynomials $P_{j-1}\left( \lambda _i\right) $ of
degrees $j-1=0,1,\ldots ,N-1$, 
\begin{equation}
\Delta \left( \lambda \right) =\det \left\| P_{j-1}\left( \lambda
_i\right) \right\| .  \label{req.6&&7}
\end{equation}
Further, choosing these polynomials to be orthogonal with respect to the
measure $d\alpha \left( \lambda \right) =\exp \left\{ -2V\left( \lambda
\right) \right\} d\lambda $, 
\begin{equation}
\int_{\lambda \in \left( -\infty ,+\infty \right) }d\alpha \left( \lambda
\right) P_n\left( \lambda \right) P_m\left( \lambda \right) =\delta _{nm},
\label{req.07}
\end{equation}
we arrive at the conclusion that the function $\Psi _0$ in Eq. (\ref{req.06b}%
) can be represented as a Slater determinant 
\begin{equation}
\Psi _0\left( \lambda _1,\ldots ,\lambda _N\right) =\frac 1{\sqrt{N!}}\det
\left\| \varphi _{j-1}\left( \lambda _i\right) \right\|   \label{req.08}
\end{equation}
that formally describes the system of $N$ fictitious non--interacting
fermions located at ``spatial'' points $\lambda _i$ and characterized by the
set of orthonormal ``eigenfunctions'' 
\begin{equation}
\varphi _n\left( \lambda \right) =P_n\left( \lambda \right) \exp \left\{
-V\left( \lambda \right) \right\} ,  \label{req.09}
\end{equation}
\begin{equation}
\int_{-\infty }^{+\infty }d\lambda \varphi _n\left( \lambda \right) \varphi
_m\left( \lambda \right) =\delta _{nm}.  \label{req.09w}
\end{equation}
Bearing in mind the representation Eq. (\ref{req.06a}), and taking into
account Eqs. (\ref{req.07}) -- (\ref{req.09}), we readily get that 
\begin{equation}
\rho \left( \lambda _1,\ldots ,\lambda _N\right) =\frac 1{N!}\det \left\|
K_N\left( \lambda _i,\lambda _j\right) \right\| _{i,j=1...N},
\label{req.10}
\end{equation}
where 
\begin{equation}
K_N\left( \lambda ,\lambda ^{\prime }\right) =\sum_{k=0}^{N-1}\varphi
_k\left( \lambda \right) \varphi _k\left( \lambda ^{\prime }\right) 
\label{req.11}
\end{equation}
is the {\it scalar two--point kernel}. Making use of the
Christoffel--Darboux theorem \cite{S-1967} (see Eq. (\ref{eq.20}) below),
the two--point kernel can be reduced to the form 
\begin{equation}
K_N\left( \lambda ,\lambda ^{\prime }\right) =c_N\frac{\varphi _N\left(
\lambda ^{\prime }\right) \varphi _{N-1}\left( \lambda \right) -\varphi
_N\left( \lambda \right) \varphi _{N-1}\left( \lambda ^{\prime }\right) }{%
\lambda ^{\prime }-\lambda },  \label{req.12}
\end{equation}
convenient for the further analysis. In Eq. (\ref{req.12}) $c_N$ is the
coefficient in the three--term recurrence equation (see Eq. (\ref{eq.16})
below) for the set $P_n$ of the polynomials orthogonal with respect to the
measure $d\alpha $. This formula simplifies significantly RMT calculations,
since the ``eigenfunctions'' $\varphi _N$ with large ``quantum numbers'' $%
N\gg 1$ entering Eq. (\ref{req.12}) can be replaced by their large--$N$
asymptotics.

The $n$--point correlation function is determined in the Random Matrix
Theory by the formula \cite{M-1991} 
\begin{equation}
R_n\left( \lambda _1,\ldots ,\lambda _n\right) =\frac{N!}{\left( N-n\right) !%
}\prod_{k=n+1}^N\int_{-\infty }^{+\infty }d\lambda _k\rho \left( \lambda
_1,\ldots ,\lambda _N\right) .  \label{req.13}
\end{equation}
It describes the probability to find $n$ levels around each of the points $%
\lambda _1,\ldots ,\lambda _n$ when the positions of the remaining levels
are unobserved. The multiple integrals in the last equation can exactly be
calculated by using the representation Eq. (\ref{req.10}). The result of the
integration reads \cite{M-1991} 
\begin{equation}
R_n\left( \lambda _1,\ldots ,\lambda _n\right) =\det \left\| K_N\left(
\lambda _i,\lambda _j\right) \right\| _{i,j=1...n}.  \label{req.14}
\end{equation}
Equation (\ref{req.14}) implies that the knowledge of the scalar two--point
kernel $K_N\left( \lambda ,\lambda ^{\prime }\right) $ is sufficient to
calculate the energy level correlation function of any order. In particular,
the averaged density of states is expressed as 
\begin{equation}
\left\langle \nu _N\left( \lambda \right) \right\rangle =R_1\left( \lambda
\right) =K_N\left( \lambda ,\lambda \right).  \label{wrmequ.11}
\end{equation}
Analogously, Eq. (\ref{req.14}) leads to the following expression for
connected ``density-density'' correlation function $\rho _c^{\left(
N\right) }=\left\langle \nu _N\left( \lambda \right) \nu _N\left( \lambda
^{\prime }\right) \right\rangle -\left\langle \nu _N\left( \lambda \right)
\right\rangle \left\langle \nu _N\left( \lambda ^{\prime }\right)
\right\rangle $, 
\begin{eqnarray}
\rho _c^{\left( N\right) } &=&\delta \left( \lambda -\lambda ^{\prime
}\right) R_1\left( \lambda \right) +R_2\left( \lambda ,\lambda ^{\prime
}\right)   \nonumber  \label{wrmequ.13} \\
&=&\delta \left( \lambda -\lambda ^{\prime }\right) K_N\left( \lambda
,\lambda \right) -K_N^2\left( \lambda ,\lambda ^{\prime }\right) .
\label{wrmequ.13}
\end{eqnarray}
Thus, all the nontrivial information about eigenlevel correlations is
contained in the squared two--point kernel $-K_N^2\left( \lambda ,\lambda
^{\prime }\right) $.

\section{The Shohat Method: General Relations}

Equations (\ref{req.06a}) and (\ref{req.08}) suggest that the joint
distribution function of $N$ eigenvalues of a $%
\mathop{\rm U}
\left( N\right) $ invariant random matrix ensemble can be interpreted as a
probability of finding $N$ fictitious non--interacting fermions to be
confined in a one--dimensional space. The effective one--particle
Schr\"{o}dinger equation for the wave functions $\varphi _n\left( \lambda
\right) $, Eq. (\ref{req.09}), of these fictitious fermions can be derived
by mapping a three--term recurrence equation for orthogonal polynomials, Eq.
(\ref{req.07}), onto a second--order differential equation. The method of
reducing a recurrence equation to a differential equation is essentially due
to J. Shohat who proved in 1939 that orthogonal polynomials associated with
exponential weights satisfy a second--order differential equation \cite
{S-1939}. Much later the Shohat method has got a further development in the
work by Bonan and Clark \cite{BC-1986}. By now, rather extended mathematical
literature exists on this subject \cite{JATs,N-1986,L-1993}.

Let as consider a set of polynomials $P_n\left( \lambda \right) $ orthogonal
on the entire real axis with respect to the measure $d\alpha \left( \lambda
\right) =\exp \left\{ -2V\left( \lambda \right) \right\} d\lambda $. If $%
V\left( \lambda \right) $ is an even function\footnote{%
For asymmetric confinement potentials the recurrence equation takes the form 
$\left( \lambda -b_n\right) P_{n-1}\left( \lambda \right) =c_nP_n\left(
\lambda \right) +c_{n-1}P_{n-2}\left( \lambda \right) .$ The additional
parameter $b_n$ can easily be incorporated into the calculational scheme.}, $%
V\left( -\lambda \right) =V\left( \lambda \right) $, this set of orthogonal
polynomials can be defined by the recurrence equation 
\begin{equation}
\lambda P_{n-1}\left( \lambda \right) =c_nP_n\left( \lambda \right)
+c_{n-1}P_{n-2}\left( \lambda \right) ,  \label{eq.16}
\end{equation}
where the coefficients $c_n$ are uniquely determined by the measure $d\alpha 
$. 

In order to derive the differential equation for the wave functions $\varphi
_n\left( \lambda \right) =P_n\left( \lambda \right) \exp \left\{ -V\left(
\lambda \right) \right\} $, we note that the following identity takes place, 
\begin{equation}
\frac{dP_n\left( \lambda \right) }{d\lambda }=A_n\left( \lambda \right)
P_{n-1}\left( \lambda \right) -B_n\left( \lambda \right) P_n\left( \lambda
\right),  \label{eq.17}
\end{equation}
with functions $A_n\left( \lambda \right) $ and $B_n\left( \lambda \right) $
to be determined from the following consideration. Since $dP_n\left( \lambda
\right) /d\lambda $ is a polynomial of the degree $n-1$, it can be
represented \cite{S-1967} through the Fourier expansion in the terms of the
kernel $Q_n\left( t,\lambda \right) =\sum_{k=0}^{n-1}P_k\left( \lambda
\right) P_k\left( t\right) $ as 
\begin{equation}
\frac{dP_n\left( \lambda \right) }{d\lambda }=\int d\alpha \left( t\right) 
\frac{dP_n\left( t\right) }{dt}Q_n\left( t,\lambda \right) .  \label{eq.18}
\end{equation}
Integrating by parts in the last equation we get that 
\begin{equation}
\frac{dP_n\left( \lambda \right) }{d\lambda }=2\int d\alpha \left( t\right)
Q_n\left( t,\lambda \right) \left( \frac{dV}{dt}-\frac{dV}{d\lambda }\right)
P_n\left( t\right).  \label{eq.19}
\end{equation}
Now, making use of the Christoffel--Darboux identity \cite{S-1967} 
\begin{equation}
Q_n\left( t,\lambda \right) =\sum_{k=0}^{n-1}P_k\left( \lambda \right)
P_k\left( t\right) =c_n\frac{P_n\left( t\right) P_{n-1}\left( \lambda
\right) -P_n\left( \lambda \right) P_{n-1}\left( t\right) }{t-\lambda },
\label{eq.20}
\end{equation}
we are led to 
\begin{eqnarray}
\frac{dP_n\left( \lambda \right) }{d\lambda } &=&2c_n\int d\alpha \left(
t\right) \frac{V^{\prime }\left( t\right) -V^{\prime }\left( \lambda \right) 
}{t-\lambda }P_n\left( t\right)   \nonumber \\
&&\times \left[ P_n\left( t\right) P_{n-1}\left( \lambda \right) -P_n\left(
\lambda \right) P_{n-1}\left( t\right) \right] .  \label{eq.21}
\end{eqnarray}
Comparison of this expression with Eq. (\ref{eq.17}) yields 
\label{rm.06}
\begin{eqnarray}
A_n\left( \lambda \right)  &=&2c_n\int d\alpha \left( t\right) \frac{%
V^{\prime }\left( t\right) -V^{\prime }\left( \lambda \right) }{t-\lambda }%
P_n^2\left( t\right) ,  \label{eq.22a} \\
B_n\left( \lambda \right)  &=&2c_n\int d\alpha \left( t\right) \frac{%
V^{\prime }\left( t\right) -V^{\prime }\left( \lambda \right) }{t-\lambda }%
P_n\left( t\right) P_{n-1}\left( t\right) .  \label{eq.22b}
\end{eqnarray}

Now one can obtain the {\it exact} differential equation for the
eigenfunctions $\varphi _n$. Differentiating Eq. (\ref{eq.17}), consequently
applying Eqs. (\ref{eq.17}) and (\ref{eq.16}), and taking into account Eq. (%
\ref{req.09}) we derive after somewhat lengthy calculations 

\begin{equation}
\frac{d^2\varphi _n\left( \lambda \right) }{d\lambda ^2}-{\cal F}_n\left(
\lambda \right) \frac{d\varphi _n\left( \lambda \right) }{d\lambda }+{\cal G}%
_n\left( \lambda \right) \varphi _n\left( \lambda \right) =0,  \label{eq.30}
\end{equation}
where 
\label{eq.31s}
\begin{eqnarray}
{\cal F}_n\left( \lambda \right)  &=&\frac 1{A_n}\frac{dA_n}{d\lambda },
\label{eq.31a} \\
{\cal G}_n\left( \lambda \right)  &=&\frac{dB_n}{d\lambda }+\frac{c_n}{%
c_{n-1}}A_nA_{n-1}-B_n\left( B_n+2\frac{dV}{d\lambda }+\frac 1{A_n}\frac{dA_n%
}{d\lambda }\right)   \nonumber  \label{eq.31b} \\
&&+\frac{d^2V}{d\lambda ^2}-\left( \frac{dV}{d\lambda }\right) ^2-\frac 1{A_n%
}\frac{dA_n}{d\lambda }\frac{dV}{d\lambda }.  \label{eq.31b}
\end{eqnarray}

When deriving Eqs. (\ref{eq.30}), (\ref{eq.31a}) and (\ref{eq.31b}) we made use of the sum
rule 
\begin{eqnarray}
B_n+B_{n-1}-\frac \lambda {c_{n-1}}A_{n-1}=-2\frac{dV}{d\lambda },
\label{eq.29}
\end{eqnarray}
directly following from Eqs. (\ref{eq.22a}), (\ref{eq.22b}), (\ref{eq.16})
and from oddness of $dV/d\lambda $.

Equation (\ref{eq.30}) is valid for arbitrary $n$. Previously, an equation
of this type was known in the context of the Random Matrix Theory only for
GUE, where $V\left( \lambda \right) =\lambda ^2/2$. For such a confinement
potential both functions $A_n\left( \lambda \right) $ and $B_n\left( \lambda
\right) $ can readily be obtained from Eqs. (\ref{eq.22a}) and (\ref{eq.22b}%
), and are given by $A_n\left( \lambda \right) =2c_n$ and $B_n\left( \lambda
\right) =0$. Taking into account that for GUE $c_n=\left( n/2\right) ^{1/2}$%
, we end up with ${\cal F}_n\left( \lambda \right) =0$ and ${\cal G}_n\left(
\lambda \right) =2n+1-\lambda ^2$. This allows us to interpret $\varphi
_n\left( \lambda \right) $ as a wave function of the fermion confined by a
parabolic potential, 
\begin{equation}
\frac{d^2\varphi _n^{\rm{GUE}}\left( \lambda \right) }{d\lambda ^2}+\left(
2n+1-\lambda ^2\right) \varphi _n^{\rm{GUE}}\left( \lambda \right) =0.
\label{eq.gue}
\end{equation}
The effective Schr\"{o}dinger equation (\ref{eq.30}) applies to general
non--Gaussian random matrix ensembles as well, although the explicit
calculation of ${\cal F}_n\left( \lambda \right) $ and ${\cal G}_n\left(
\lambda \right) $ in this situation is a rather complicated task. In two
cases of relatively simple measures with $V\left( \lambda \right) =$ $%
\lambda ^4/8+q_3\lambda ^3/6+q_2\lambda ^2/4+q_1\lambda /2$ and $V\left(
\lambda \right) =\lambda ^6/12$ the functions ${\cal F}_n\left( \lambda
\right) $ and ${\cal G}_n\left( \lambda \right) $ are known explicitly \cite
{JATs}. Significant simplifications, however, arise in the limit $n=N\gg 1$,
which is just a thermodynamic limit of the Random Matrix Theory representing
for us the most interest.

\section{Random Matrices with Single Eigenvalue Support}

In this Section we will be interested in the study of eigenvalue statistics
for unitary invariant non--Gaussian large random matrices characterized by a
distribution function $P\left[ {\bf H}\right] $ given by Eq. (\ref{1s}). The
confinement potential associated with this model is 
\begin{equation}
V_\alpha \left( \lambda \right) =v\left( \lambda \right) -\alpha \log \left|
\lambda \right| .  \label{v}
\end{equation}
Here $v\left( \lambda \right) $ is the regular part of level confinement 
\begin{equation}
v\left( \lambda \right) =\sum_{k=1}^p\frac{d_k}{2k}\lambda ^{2k},  \label{vr}
\end{equation}
with a positive leading coefficient, $d_p>0$; the signs of the rest of the $%
d_k$ can be arbitrary but they should lead to an eigenvalue density
supported on a single connected interval, $\left\{ \lambda \right\} \in
\left( -{\cal D}_N,+{\cal D}_N\right) $. The parameter $\alpha >-1/2$ is the
strength of the logarithmic singularity.

In Subsection 4.1 below, we demonstrate how the one--point spectral
characteristics (density of states) can be obtained by making use of the
recurrence equation (\ref{eq.16}). In Subsection 4.2, we turn to the study
of the smoothed connected ``density--density'' correlator, also starting
with recurrence equation (\ref{eq.16}). Finally, in Subsection 4.3, we
obtain the universal scalar kernels in the origin, bulk and soft--edge
scaling limits by solving the effective Schr\"{o}dinger equation for
fictitious fermions.

\subsection{Macroscopic Level Density from Recurrence Equation}

We start with an explanation of the main idea of the derivation to make
clear all the subsequent calculations. Here we mainly follow Ref. \cite
{AK-1998}. Our basic observation is that in the large--$N$ limit the density
of states\footnote{%
Hereafter we use the notation $\nu _N\left( \lambda \right) $ for averaged density of states, Eq. (\ref{wrmequ.11}).} $\nu _N\left( \lambda \right) $ consists of two parts, 
\begin{equation}
\nu _N\left( \lambda \right) =\nu _N^{\rm{smooth}}\left( \lambda \right)
+\nu _N^{\rm{osc}}\left( \lambda \right) .  \label{akeq.04}
\end{equation}
The smooth part $\nu _N^{\rm{smooth}}\left( \lambda \right) $ contributes
to different integral characteristics determined by the density of states,
while the oscillating part does not, because any integration will level the
oscillating features. Then, for some smooth, well behaved, even function $%
f\left( \lambda \right) $ we have 
\begin{equation}
\int_{-\infty }^{+\infty }d\lambda f\left( \lambda \right) \nu _N\left(
\lambda \right) \stackrel{N\rightarrow \infty }{\rightarrow }\int_{\lambda
\in \rm{support}}d\lambda f\left( \lambda \right) \nu _N^{\rm{smooth}%
}\left( \lambda \right) .  \label{akeq.05}
\end{equation}

Let us implement this scheme by choosing (without any loss of generality) $%
f\left( \lambda \right) =\lambda ^{2s}$, with $s$ being a positive integer.
This choice is possible due to the evenness of $\nu _N\left( \lambda \right) 
$. By definition Eq. (\ref{wrmequ.11}), we obtain from Eq. (\ref{req.12}) 
\begin{equation}
\nu _N\left( \lambda \right) =c_N\exp \left\{ -2V_\alpha \left( \lambda
\right) \right\} \left( P_N^{\left( \alpha \right) \prime }\left( \lambda
\right) P_{N-1}^{\left( \alpha \right) }\left( \lambda \right)
-P_{N-1}^{\left( \alpha \right) \prime }\left( \lambda \right) P_N^{\left(
\alpha \right) }\left( \lambda \right) \right) .  \label{akeq.01}
\end{equation}
Having in mind the relation Eq. (\ref{eq.17}), the sum rule Eq. (\ref{eq.29}%
) and the definition Eq. (\ref{req.09}) we come down to 

\begin{eqnarray}
\nu _N\left( \lambda \right)  &=&c_N\left[ A_N^{\left( \alpha \right)
}\left( \lambda \right) \left( \varphi _{N-1}^{\left( \alpha \right) }\left(
\lambda \right) \right) ^2+\frac{c_N}{c_{N-1}}A_{N-1}^{\left( \alpha \right)
}\left( \lambda \right) \left( \varphi _N^{\left( \alpha \right) }\left(
\lambda \right) \right) ^2\right.   \label{akeq.03} \\
&&\left. -\varphi _N^{\left( \alpha \right) }\left( \lambda \right) \varphi
_{N-1}^{\left( \alpha \right) }\left( \lambda \right) \left( \frac \lambda {%
c_{N-1}}A_{N-1}^{\left( \alpha \right) }\left( \lambda \right) +B_N^{\left(
\alpha \right) }\left( \lambda \right) -B_{N-1}^{\left( \alpha \right)
}\left( \lambda \right) \right) \right] .  \nonumber  \label{akeq.0345}
\end{eqnarray}
Here the upper index $\left( \alpha \right) $ is used to reflect the
presence of a log--singular component in confinement potential $V_\alpha
\left( \lambda \right) $, Eq. (\ref{v}). Thus, the level density is
expressed through the wave functions $\varphi _N^{\left( \alpha \right)
}\left( \lambda \right) $, and through the functions $A_N^{\left( \alpha
\right) }\left( \lambda \right) $ and $B_N^{\left( \alpha \right) }\left(
\lambda \right) $, given by Eqs. (\ref{eq.22a}) and (\ref{eq.22b}).

For further convenience we introduce two quantities, 
\begin{equation}
\Lambda _{2\sigma }^{\left( N\right) }=\int d\alpha \left( t\right) \left(
P_N^{\left( \alpha \right) }\left( t\right) \right) ^2t^{2\sigma
}=\int_{-\infty }^{+\infty }dt\left( \varphi _N^{\left( \alpha \right)
}\left( t\right) \right) ^2t^{2\sigma },  \label{akeq.06}
\end{equation}
\begin{equation}
\Gamma _{2\sigma +1}^{\left( N\right) }=\int d\alpha \left( t\right)
P_N^{\left( \alpha \right) }\left( t\right) P_{N-1}^{\left( \alpha \right)
}\left( t\right) t^{2\sigma +1}=\int_{-\infty }^{+\infty }dt\varphi
_N^{\left( \alpha \right) }\left( t\right) \varphi _{N-1}^{\left( \alpha
\right) }\left( t\right) t^{2\sigma +1},  \label{akeq.07}
\end{equation}
for which the alternative explicit integral representations are shown to
exist in the Appendix A. Without going into details of those calculations,
we only stress an extremely important role played by both the large--$N$
limit, Eq. (\ref{a2}), of the recurrence equation for associated orthogonal
polynomials and the asymptotic expansion, Eq. (\ref{a3}), deduced from Eq. (\ref{a2}).
We also notice that owing to the existence of these explicit
representations, derived by using a large--$N$ version of the recurrence
equation (\ref{eq.16}), all further calculations became possible.

The functions $A_N^{\left( \alpha \right) }$ and $B_N^{\left( \alpha \right)
}$ entering Eq. (\ref{akeq.03}) can be expressed in terms of $\Lambda
_\sigma ^{\left( N\right) }$ and $\Gamma _\sigma ^{\left( N\right) }$.
Namely, bearing in mind the definitions given by Eqs. (\ref{eq.22a}) and (%
\ref{eq.22b}) we obtain for $A_N^{\left( \alpha \right) }\left( \lambda
\right) =A_{\rm{reg}}^{\left( N\right) }\left( \lambda \right) +\alpha A_{%
\rm{sing}}^{\left( N\right) }\left( \lambda \right) $, 
\begin{eqnarray}
A_{\rm{reg}}^{\left( N\right) }\left( \lambda \right)
&=&2c_N\sum_{k=1}^pd_k\sum_{\sigma =1}^k\Lambda _{2\left( k-\sigma \right)
}^{\left( N\right) }\lambda ^{2\sigma -2},  \label{eq.a02} \\
A_{\rm{sing}}^{\left( N\right) }\left( \lambda \right) &=&2c_N\int \frac{%
d\alpha \left( t\right) }t\left( P_N^{\left( \alpha \right) }\left( t\right)
\right) ^2.  \label{eq.a03}
\end{eqnarray}
In the same way, the function $B_N^{\left( \alpha \right) }\left( \lambda
\right) =B_{\rm{reg}}^{\left( N\right) }\left( \lambda \right) +\alpha B_{%
\rm{sing}}^{\left( N\right) }\left( \lambda \right) $ is given by 
\begin{eqnarray}
B_{\rm{reg}}^{\left( N\right) }\left( \lambda \right)
&=&2c_N\sum_{k=1}^pd_k\sum_{\sigma =1}^k\Gamma _{2\left( k-\sigma \right)
-1}^{\left( N\right) }\lambda ^{2\sigma -1},  \label{eq.b02} \\
B_{\rm{sing}}^{\left( N\right) }\left( \lambda \right) &=&\frac{2c_N}%
\lambda \int \frac{d\alpha \left( t\right) }tP_N^{\left( \alpha \right)
}\left( t\right) P_{N-1}^{\left( \alpha \right) }\left( t\right) .
\label{eq.b03}
\end{eqnarray}
When deriving these formulas we have used the fact of evenness of the
measure $d\alpha \left( t\right) /dt$ and of $\left( P_N^{\left( \alpha
\right) }\left( t\right) \right) ^2$, as well as the expansion 
\begin{equation}
\frac{t^k-\lambda ^k}{t-\lambda }=\sum_{m=1}^kt^{k-m}\lambda ^{m-1}.
\label{exp.exp}
\end{equation}

The ``singular'' components, $A_{\rm{sing}}^{\left( N\right) }$ and $B_{%
\rm{sing}}^{\left( N\right) }$, can easily be determined. First, due to
oddness of the integrand in Eq. (\ref{eq.a03}), we have $A_{\rm{sing}%
}^{\left( N\right) }\left( \lambda \right) \equiv 0$. Second, in order to
find $B_{\rm{sing}}^{\left( N\right) }$, we notice that the quantity 
\begin{equation}
\gamma _n^{\left( \alpha \right) }=c_n\int \frac{d\alpha \left( t\right) }t%
P_n^{\left( \alpha \right) }\left( t\right) P_{n-1}^{\left( \alpha \right)
}\left( t\right) ,  \label{dos.45}
\end{equation}
where $n$ is not necessarily large, obeys the identity $\gamma _n^{\left(
\alpha \right) }+\gamma _{n-1}^{\left( \alpha \right) }=1,$ which is a
direct consequence of the recurrence equation (\ref{eq.16}). As far as $%
\gamma _{2n}^{\left( \alpha \right) }=\gamma _{2n-2}^{\left( \alpha \right)
}=\ldots =\gamma _2^{\left( \alpha \right) }\equiv 0,$ we conclude that $%
\gamma _n^{\left( \alpha \right) }=\left[ 1-\left( -1\right) ^n\right] /2$
and therefore, $B_{\rm{sing}}^{\left( N\right) }\left( \lambda \right) =$ $%
2\gamma _N^{\left( \alpha \right) }/\lambda $. Hence we are led to the
following representations, 
\begin{eqnarray}
A_N^{\left( \alpha \right) }\left( \lambda \right)
&=&2c_N\sum_{k=1}^pd_k\sum_{\sigma =1}^k\Lambda _{2\left( k-\sigma \right)
}^{\left( N\right) }\lambda ^{2\sigma -2},  \label{eq.A} \\
B_N^{\left( \alpha \right) }\left( \lambda \right)
&=&2c_N\sum_{k=1}^pd_k\sum_{\sigma =1}^k\Gamma _{2\left( k-\sigma \right)
-1}^{\left( N\right) }\lambda ^{2\sigma -1}+\alpha \frac{1-\left( -1\right)
^N}\lambda .  \label{eq.B}
\end{eqnarray}

With these preliminarily calculations in hand we are ready to implement the
idea of recovering the Dyson density of states from the recurrence equation.
In accordance with Eqs. (\ref{akeq.05}) and (\ref{akeq.03}) there are five
contributions to the integral in the r.h.s. of Eq. (\ref{akeq.05})
corresponding to five terms in Eq. (\ref{akeq.03}). Substituting Eqs. (\ref
{eq.A}) and (\ref{eq.B}) into Eq. (\ref{akeq.03}) and performing a formal
integration with the help of Eqs. (\ref{akeq.06}) and (\ref{akeq.07}), these
contributions are found to be 
\label{peq.31}
\begin{eqnarray}
\rho _1 &=&2c_N^2\sum_{k=1}^pd_k\sum_{\sigma =1}^k\Lambda _{2\left( k-\sigma
\right) }^{\left( N\right) }\Lambda _{2\left( \sigma +s-1\right) }^{\left(
N-1\right) },  \label{r1} \\
\rho _2 &=&2c_N^2\sum_{k=1}^pd_k\sum_{\sigma =1}^k\Lambda _{2\left( k-\sigma
\right) }^{\left( N-1\right) }\Lambda _{2\left( \sigma +s-1\right) }^{\left(
N\right) },  \label{r2} \\
\rho _3 &=&-2c_N\sum_{k=1}^pd_k\sum_{\sigma =1}^k\Lambda _{2\left( k-\sigma
\right) }^{\left( N-1\right) }\Gamma _{2\left( \sigma +s\right) -1}^{\left(
N\right) },  \label{r3} \\
\rho _4 &=&-2c_N^2\sum_{k=1}^pd_k\sum_{\sigma =1}^{k-1}\Gamma _{2\left(
k-\sigma \right) -1}^{\left( N\right) }\Gamma _{2\left( \sigma +s\right)
-1}^{\left( N\right) }  \nonumber  \label{r4} \\
&&-\alpha c_N\left[ 1-\left( -1\right) ^N\right] \Gamma _{2s-1}^{\left(
N\right) },  \label{r4} \\
\rho _5 &=&2c_Nc_{N-1}\sum_{k=1}^pd_k\sum_{\sigma =1}^{k-1}\Gamma _{2\left(
k-\sigma \right) -1}^{\left( N-1\right) }\Gamma _{2\left( \sigma +s\right)
-1}^{\left( N\right) }  \nonumber  \label{r5} \\
&&+\alpha c_N\left[ 1-\left( -1\right) ^{N-1}\right] \Gamma _{2s-1}^{\left(
N\right) }.  \label{r5}
\end{eqnarray}
For large--$N$ matrix models with a single spectrum support there are
asymptotic identities $c_N\approx c_{N-1}$, $\Lambda _\sigma ^{\left(
N\right) }\approx \Lambda _\sigma ^{\left( N-1\right) }$ and $\Gamma _\sigma
^{\left( N\right) }\approx \Gamma _\sigma ^{\left( N-1\right) }$ (see
Appendix A for details) which simplify matters greatly. Collecting Eqs. (\ref
{r1}) -- (\ref{r5}) we come down to 
\begin{eqnarray}
&&\int_{-\infty }^{+\infty }d\lambda \lambda ^{2s}\nu _N\left( \lambda
\right) \stackrel{N\rightarrow \infty }{\rightarrow }-2\alpha c_N\left(
-1\right) ^N\Gamma _{2s-1}^{\left( N\right) }  \nonumber \\
+ &&2c_N\sum_{k=1}^pd_k\sum_{\sigma =1}^k\Lambda _{2\left( k-\sigma \right)
}^{\left( N\right) }\left( \Lambda _{2\left( \sigma +s-1\right) }^{\left(
N\right) }-2c_N\Gamma _{2\left( \sigma +s\right) -1}^{\left( N\right)
}\right) .  \label{rr}
\end{eqnarray}
Further double summation over indices $k$ and $\sigma $ is performed
by making use of the integral representations for $\Lambda _\sigma
^{\left( N\right) }$ and $\Gamma _\sigma ^{\left( N\right) }$ given by Eqs.
(\ref{a7}) and (\ref{a9}) in Appendix A. Straightforward calculations lead to 
\begin{equation}
\int_{-\infty }^{+\infty }d\lambda \lambda ^{2s}\nu _N\left( \lambda \right) 
\stackrel{N\rightarrow \infty }{\rightarrow }\int_{-{\cal D}_N}^{{\cal D}%
_N}d\lambda \lambda ^{2s}\rho _\Sigma \left( \lambda \right) ,
\label{akeq.19a}
\end{equation}
with ${\cal D}_N=2c_N$ and 
\begin{eqnarray}
\rho _\Sigma \left( \lambda \right) &=&\frac 2{\pi ^2}\left( {\cal D}%
_N^2-\lambda ^2\right) ^{1/2}{\cal P}\int_0^{{\cal D}_N}\frac{dt}{\left( 
{\cal D}_N^2-t^2\right) ^{1/2}}\frac{tv^{\prime }\left( t\right) -\lambda
v^{\prime }\left( \lambda \right) }{t^2-\lambda ^2}  \nonumber
\label{akeq.19b} \\
&&-\frac \alpha \pi \frac{\left( -1\right) ^N}{\left( {\cal D}_N^2-\lambda
^2\right) ^{1/2}}.  \label{akeq.19b}
\end{eqnarray}
For $\lambda ={\cal D}_Nz$ with $\left| z\right| <1$ the term proportional
to $\lambda dv/d\lambda $ vanishes identically due to the principal value $%
{\cal P}$ of the integral over variable $t$, while the term proportional to $%
\alpha $ is a subleading one in the large--$N$ limit. Then, in accordance
with our concept Eq. (\ref{akeq.05}), we end up with 
\begin{equation}
\nu _N^{\rm{smooth}}\left( \lambda \right) =\frac 2{\pi ^2}\left( {\cal D}%
_N^2-\lambda ^2\right) ^{1/2}{\cal P}\int_0^{{\cal D}_N}\frac{tdt}{\left( 
{\cal D}_N^2-t^2\right) ^{1/2}}\frac{dv/dt}{t^2-\lambda ^2}.  \label{akeq.18}
\end{equation}
Equation (\ref{akeq.18}) is exactly the Dyson density $\nu _D\left( \lambda
\right) $ of states with ${\cal D}_N$ being the end point of the eigenvalue
support. We reconstructed the macroscopic level density Eq. (\ref{akeq.18})
directly from the recurrence equation (\ref{eq.16}), alternatively to the
traditional mean--field--theory derivation \cite{M-1991}. Notice that the
spectrum end point ${\cal D}_N$ is the positive root to the integral
equation 
\begin{equation}
\frac{\pi N}2=\int_0^{{\cal D}_N}\frac{tdt}{\left( {\cal D}_N^2-t^2\right)
^{1/2}}\frac{dv}{dt}  \label{mrs}
\end{equation}
following from the normalization of the level density.

\subsection{Global Connected ``Density--Density'' Correlation Function}

The same technology is applicable to the study of the smoothed connected
``density--density'' correlator $\rho _c^{\left( N\right) }$. It is defined
in terms of the scalar kernel by Eq. (\ref{wrmequ.13}), so that 
\begin{eqnarray}
\rho _c^{\left( N\right) }\left( \lambda ,\lambda ^{\prime }\right)  &=&-%
\overline{K_N^2\left( \lambda ,\lambda ^{\prime }\right) }=-\frac{c_N^2}{%
\left( \lambda -\lambda ^{\prime }\right) ^2}  \nonumber \\
&&\times \left[ \overline{\left( \varphi _N^{\left( \alpha \right) }\left(
\lambda \right) \right) ^2}\,\overline{\left( \varphi _{N-1}^{\left( \alpha
\right) }\left( \lambda ^{\prime }\right) \right) ^2}+\overline{\left(
\varphi _{N-1}^{\left( \alpha \right) }\left( \lambda \right) \right) ^2}\,%
\overline{\left( \varphi _N^{\left( \alpha \right) }\left( \lambda ^{\prime
}\right) \right) ^2}\right.   \nonumber \\
&&\left. -2\overline{\varphi _N^{\left( \alpha \right) }\left( \lambda
\right) \varphi _{N-1}^{\left( \alpha \right) }\left( \lambda \right) }\,%
\overline{\varphi _N^{\left( \alpha \right) }\left( \lambda ^{\prime
}\right) \varphi _{N-1}^{\left( \alpha \right) }\left( \lambda ^{\prime
}\right) }\right] .
\begin{array}{c}
\\ 
\end{array}
\label{sc.01}
\end{eqnarray}
We remind that here $\varphi _N^{\left( \alpha \right) }\left( \lambda
\right) =\exp \left\{ -V_\alpha \left( \lambda \right) \right\} P_N^{\left(
\alpha \right) }\left( \lambda \right) $ are fictitious wave functions$,$ $%
\lambda \neq \lambda ^{\prime }$, and $\overline{\left( \ldots \right) }$
denotes averaging over rapid oscillations manifested on the characteristic
scale of the mean level spacing. The averaging in Eq. (\ref{sc.01}) can be
done along the lines of the previous Subsection with two modifications.
First, as far as $\lambda \neq \lambda ^{\prime }$ we can run averaging over 
$\lambda $ and $\lambda ^{\prime }$ independently (this is already reflected
in Eq. (\ref{sc.01})). Second, we have to take into account the evenness of $%
\left( \varphi _N^{\left( \alpha \right) }\left( \lambda \right) \right) ^2$
and the oddness of $\varphi _N^{\left( \alpha \right) }\left( \lambda \right)
\varphi _{N-1}^{\left( \alpha \right) }\left( \lambda \right) $.

There are two integrals 
\label{sc.02}
\begin{eqnarray}
I_1^{\left( N\right) } &=&\int_{-\infty }^{+\infty }d\lambda \lambda
^{2s}\exp \left\{ -2V_\alpha \left( \lambda \right) \right\} \left(
P_{N-1}^{\left( \alpha \right) }\left( \lambda \right) \right) ^2,
\label{sc.02a} \\
I_2^{\left( N\right) } &=&\int_{-\infty }^{+\infty }d\lambda \lambda
^{2s+1}\exp \left\{ -2V_\alpha \left( \lambda \right) \right\} P_N^{\left(
\alpha \right) }\left( \lambda \right) P_{N-1}^{\left( \alpha \right)
}\left( \lambda \right)   \label{sc.02b}
\end{eqnarray}
to be evaluated in the large--$N$ limit. With the help of Eqs. (\ref{akeq.06}%
) and (\ref{akeq.07}) one immediately recognizes them as the objects $%
\Lambda _{2s}^{\left( N\right) }$ and $\Gamma _{2s+1}^{\left( N\right) }$,
respectively, calculated in Appendix A. By comparing of Eqs. (\ref{sc.02a})
and (\ref{sc.02b}) with Eqs. (\ref{a7}) and (\ref{a9}) we deduce that 
\begin{eqnarray}
\overline{\left( \varphi _N^{\left( \alpha \right) }\left( \lambda \right)
\right) ^2} &=&\frac 1\pi \left( {\cal D}_N^2-\lambda ^2\right) ^{-1/2},
\label{sc.03a} \\
\overline{\varphi _N^{\left( \alpha \right) }\left( \lambda \right) \varphi
_{N-1}^{\left( \alpha \right) }\left( \lambda \right) } &=&\frac \lambda {%
\pi {\cal D}_N}\left( {\cal D}_N^2-\lambda ^2\right) ^{-1/2},  \label{sc.03b}
\end{eqnarray}
for $\left| \lambda \right| <{\cal D}_N$. Substituting Eqs. (\ref{sc.03a}) and (%
\ref{sc.03b}) into Eq. (\ref{sc.01}) is a final step leading us to the
smoothed ``density--density'' correlator 
\begin{equation}
\rho _c^{\left( N\right) }\left( \lambda ,\lambda ^{\prime }\right) =-\frac 1%
{2\pi ^2\left( \lambda -\lambda ^{\prime }\right) ^2}\frac{{\cal D}%
_N^2-\lambda \lambda ^{\prime }}{\left( {\cal D}_N^2-\lambda ^2\right)
^{1/2}\left( {\cal D}_N^2-\lambda ^{\prime 2}\right) ^{1/2}}  \label{sc.04}
\end{equation}
announced by Eq. (\ref{gu}) with $\beta =2$. We stress, that it has been
obtained here from the recurrence equation for orthogonal polynomials
associated with the random matrix ensemble in question.

\subsection{Local Eigenvalue Correlations by Solving Effective
Schr\"{o}dinger Equation}

\subsubsection{Effective Schr\"{o}dinger Equation}

Local eigenvalue correlations in the matrix ensemble Eq. (\ref{1s}) can be
studied by using the asymptotic version of the effective Schr\"{o}dinger
equation (\ref{eq.30}) obtained in Section 3. For the confinement
potential $V_\alpha $ introduced by Eq. (\ref{v}) we obtain in the large--$N$
limit the following expressions for the functions ${\cal F}_n^{\left( \alpha
\right) }\left( \lambda \right) $ and ${\cal G}_n^{\left( \alpha \right)
}\left( \lambda \right) $ (see Eqs. (\ref{eq.31a}) and (\ref{eq.31b})), 
\begin{eqnarray}
{\cal F}_N^{\left( \alpha \right) }\left( \lambda \right)  &=&\frac d{%
d\lambda }\log A_N^{\left( \alpha \right) }\left( \lambda \right) ,
\label{pm.01a} \\
{\cal G}_N^{\left( \alpha \right) }\left( \lambda \right)  &=&\left(
A_N^{\left( \alpha \right) }\left( \lambda \right) \right) ^2\left[ 1-\left( 
\frac \lambda {{\cal D}_N}\right) ^2\right]   \nonumber  \label{pm.01b} \\
&&+\left( -1\right) ^N\frac \alpha \lambda \left( \frac d{d\lambda }\log
A_N^{\left( \alpha \right) }\left( \lambda \right) \right) +\frac{\alpha
\left( -1\right) ^N-\alpha ^2}{\lambda ^2}.  \label{pm.01b}
\end{eqnarray}
Here we have used the sum rule Eq. (\ref{eq.29}) to eliminate $B_N^{\left(
\alpha \right) }$. For confinement potential with a smooth regular part $v$,
the second term in Eq. (\ref{pm.01b}) is a subleading and hence it must be
discarded\footnote{%
This is not the case for the {\it multicritical} correlations near the origin $%
\lambda =0$. A detailed discussion of this important situation can be found in the very recent paper of Ref. \cite
{ADMN-1998}.}. Note, that it is rather interesting that the confinement
potential $V_\alpha $ does not appear in both equations above in an explicit
way. It is even more exciting that in the considered approximation the
function $A_N^{\left( \alpha \right) }$ can be solely expressed through the
Dyson density, Eq. (\ref{akeq.18}). Indeed, taking into account the
representation Eq. (\ref{a10}) and the fact that $A_N^{\left( \alpha \right)
}\left( \lambda \right) =A_{\rm{reg}}^{\left( N\right) }\left( \lambda
\right) $, we are led to the asymptotic relation 
\begin{equation}
A_N^{\left( \alpha \right) }\left( \lambda \right) =\frac{\pi \nu _D\left(
\lambda \right) }{\left[ 1-\left( \lambda /{\cal D}_N\right) ^2\right] ^{1/2}%
},  \label{pm.02}
\end{equation}
where, in accordance with Eq. (\ref{akeq.18}), 
\begin{equation}
\nu _D\left( \lambda \right) =\frac 2{\pi ^2}{\cal P}\int_0^{{\cal D}_N}%
\frac{tdt}{t^2-\lambda ^2}\frac{dv}{dt}\left( \frac{{\cal D}_N^2-\lambda ^2}{%
{\cal D}_N^2-t^2}\right) ^{1/2}.  \label{pm.025}
\end{equation}
This allows us to arrive at the following remarkable effective one--particle
Schr\"{o}dinger equation for the wave functions 
\begin{eqnarray}
\varphi _N^{\left( \alpha
\right) }\left( \lambda \right) =\left| \lambda \right| ^\alpha P_N^{\left(
\alpha \right) }\left( \lambda \right) \exp \left\{ -v\left( \lambda \right)
\right\} \label{wf}
\end{eqnarray}
of fictitious non--interacting fermions in the large--$N$ limit 
\cite{KF-1998,KF-1997a}
\begin{eqnarray}
\frac{d^2\varphi _N^{\left( \alpha \right) }}{d\lambda ^2}&- &\left[ \frac d{%
d\lambda }\log \left( \frac{\pi \nu _D\left( \lambda \right) }{\left[
1-\left( \lambda /{\cal D}_N\right) ^2\right] ^{1/2}}\right) \right] \frac{%
d\varphi _N^{\left( \alpha \right) }}{d\lambda }  \nonumber  \label{pm.03} \\
&+ &\left( \pi ^2\nu _D^2\left( \lambda \right) +\frac{\left( -1\right)
^N\alpha -\alpha ^2}{\lambda ^2}\right) \varphi _N^{\left( \alpha \right)
}\left( \lambda \right) =0.  \label{pm.03}
\end{eqnarray}
Also, due to Eq. (\ref{eq.17}), one can verify that the wave functions $%
\varphi _{N-1}^{\left( \alpha \right) }\left( \lambda \right) $ and $\varphi
_N^{\left( \alpha \right) }\left( \lambda \right) $ of two successive
quantum states are connected by the relationship 
\begin{equation}
\frac{d\varphi _N^{\left( \alpha \right) }}{d\lambda }=\frac{\pi \nu
_D\left( \lambda \right) }{\left[ 1-\left( \lambda /{\cal D}_N\right)
^2\right] ^{1/2}}\left( \varphi _{N-1}^{\left( \alpha \right) }\left(
\lambda \right) -\frac \lambda {{\cal D}_N}\varphi _N^{\left( \alpha \right)
}\left( \lambda \right) \right) +\left( -1\right) ^N\frac \alpha \lambda
\varphi _N^{\left( \alpha \right) }\left( \lambda \right) .  \label{pm.04}
\end{equation}
Equations (\ref{pm.03}) and (\ref{pm.04}) serve as a general basis for the
study of eigenvalue correlations in non--Gaussian random matrix ensembles in
an {\it arbitrary spectral range}.

It is instructive to analyze them in the particular case of GUE, where the
Dyson density of states is the celebrated semicircle, 
\begin{eqnarray}
\nu _D^{\rm{{GUE}}%
}\left( \lambda \right) =\pi ^{-1}\left( {\cal D}_N^2-\lambda ^2\right)
^{1/2} \label{semi}
\end{eqnarray}
with ${\cal D}_N=\left( 2N\right) ^{1/2}$. The square--root law for $%
\nu _D^{\rm{{GUE}}}\left( \lambda \right) $ immediately removes the first
derivative $d\varphi _N^{\left( \alpha \right) }/d\lambda $ in Eq. (\ref
{pm.03}), providing us with the possibility to interpret the fictitious
fermions as those confined by a quadratic potential $\left( \alpha =0\right) 
$. As far as the semicircle is a distinctive feature of density of states in
GUE only, one will always obtain a first derivative in the effective
Schr\"{o}dinger equation for non--Gaussian unitary ensembles of random
matrices. Therefore, fictitious non--interacting fermions associated with
non--Gaussian ensembles of random matrices occur in a non--Hermitean quantum
mechanics.

An interesting property of these equations is that they do not contain the
regular part of confinement potential explicitly, but only involve the Dyson
density $\nu _D$ (analytically continued on the entire real axis) and the
spectrum end point ${\cal D}_N$. In contrast, the logarithmic singularity
(that does not affect the Dyson density) introduces additional singular
terms into Eqs. (\ref{pm.03}) and (\ref{pm.04}), changing significantly the
behavior of the wave function $\varphi _N^{\left( \alpha \right) }$ near the
origin $\lambda =0$. The influence of the singularity decreases rather
rapidly outward from the origin.

Structure of the effective Schr\"{o}dinger equation leads us to the
following statements \cite{KF-1998} valid in the thermodynamic limit:

\begin{quotation}
$\bullet $ {\it Eigenvalue correlations are stable with respect to
non--singular deformations of the confinement potential.}
\end{quotation}
\begin{quotation}
$\bullet $ In the random matrix ensembles with well behaved confinement
potential {\it the knowledge of Dyson density} (that is rather crude
one--point characteristics coinciding with the real density of states only
in the spectrum bulk) {\it is sufficient to determine the genuine density of
states, as well as the }$n$--{\it point correlation function, everywhere.}
\end{quotation}

The latter conclusion is rather unexpected since it considerably reduces the
knowledge required for computing $n$--point correlators.

Effective Schr\"{o}dinger equation obtained above enables us to examine in a
unified way the local eigenvalue correlations in non--Gaussian ensembles
with $%
\mathop{\rm U}
\left( N\right) $ symmetry in different scaling limits. As we show below, it
inevitably leads to the universal Bessel correlations in the origin scaling
limit \cite{N-1996,ADMN-1997,KF-1997a}, to the universal sine correlations
in the bulk scaling limit \cite{BZ-1993,HW-1995,FKY-1996a}, and to the
universal Airy {c}orrelations in the soft--edge scaling limit \cite{KF-1997a}%
. Corresponding scalar kernels are given by Eqs. (\ref{bkpm}), (\ref{skpm})
and (\ref{pm.18}), respectively.

\subsubsection{Origin scaling limit and the universal Bessel law}

Origin scaling limit deals with the region of the spectrum close to $\lambda
=0$ where confinement potential displays the logarithmic singularity. In the
vicinity of the origin the Dyson density can be taken as being approximately
a constant, $\nu _D\left( 0\right) =1/\Delta _N\left( 0\right) $, where $%
\Delta _N\left( 0\right) $ is the mean level spacing at the origin in the
absence of the logarithmic deformation of potential $v$. Within the
framework of this approximation, Eq. (\ref{pm.03}) reads 
\begin{equation}
\frac{d^2\varphi _N^{\left( \alpha \right) }}{d\lambda ^2}+\left( \frac{\pi
^2}{\Delta _N^2\left( 0\right) }+\frac{\left( -1\right) ^N\alpha -\alpha ^2}{%
\lambda ^2}\right) \varphi _N^{\left( \alpha \right) }\left( \lambda \right)
=0.  \label{pm.05}
\end{equation}
Solution to this equation that remains finite at $\lambda =0$ can be
expressed by means of Bessel functions 
\label{bessel}
\begin{eqnarray}
\varphi _{2N}^{\left( \alpha \right) }\left( \lambda \right) &=&a\sqrt{%
\lambda }J_{\alpha - 1/2}\left( \frac{\pi \lambda }{\Delta \left(
0\right) }\right) ,  \label{pm.06a} \\
\varphi _{2N+1}^{\left( \alpha \right) }\left( \lambda \right) &=&b\sqrt{%
\lambda }J_{\alpha + 1/2}\left( \frac{\pi \lambda }{\Delta \left(
0\right) }\right) ,  \label{pm.06b}
\end{eqnarray}
where $a$ and $b$ are constants to be determined later, and $\Delta \left(
0\right) =\Delta _{2N}\left( 0\right) \approx \Delta _{2N+1}\left( 0\right) $%
. Inserting these solutions into Eq. (\ref{req.12}) we find that the scalar
kernel can be written down as 
\begin{eqnarray}
K_{2N}^{\left( \alpha \right) }\left( \lambda ,\lambda ^{\prime }\right) &=&c%
\frac{\sqrt{\lambda \lambda ^{\prime }}}{\lambda ^{\prime }-\lambda }\left[
J_{\alpha + 1/2}\left( \frac{\pi \lambda }{\Delta \left( 0\right) }%
\right) J_{\alpha - 1/2}\left( \frac{\pi \lambda ^{\prime }}{\Delta
\left( 0\right) }\right) \right.  \nonumber \\
&&\left. -J_{\alpha + 1/2}\left( \frac{\pi \lambda ^{\prime }}{\Delta
\left( 0\right) }\right) J_{\alpha - 1/2}\left( \frac{\pi \lambda }{%
\Delta \left( 0\right) }\right) \right] ,  \label{pm.07}
\end{eqnarray}
where the unknown factor $c$ can be found from the requirement $%
K_{2N}^{\left( \alpha =0\right) }\left( \lambda ,\lambda \right)$ $=$ $1/\Delta
\left( 0\right) $. This immediately yields us the value $c=-\pi /\Delta
\left( 0\right) $. Defining now the scaled variable $s=\lambda _s/\Delta
\left( 0\right) $, we obtain that in the origin scaling limit the scalar
kernel $K_{\rm{orig}}\left( s,s^{\prime }\right) =\lim_{N\rightarrow
\infty }\lambda _s^{\prime }K_{2N}^{\left( \alpha \right) }\left( \lambda
_s,\lambda _{s^{\prime }}\right) $ takes the universal Bessel law, 
\begin{equation}
K_{\rm{orig}}\left( s,s^{\prime }\right) =\frac \pi 2\left( ss^{\prime
}\right) ^{1/2}\frac{J_{\alpha +1/2}\left( \pi s\right) J_{\alpha
-1/2}\left( \pi s^{\prime }\right) -J_{\alpha -1/2}\left( \pi s\right)
J_{\alpha +1/2}\left( \pi s^{\prime }\right) }{s-s^{\prime }}.  \label{bkpm}
\end{equation}
Equation (\ref{bkpm}) is valid for arbitrary $\alpha >-1/2$, thus extending
a recent proof \cite{ADMN-1997} of universality of the Bessel kernel.

\subsubsection{Bulk scaling limit and the universal sine law}

Bulk scaling limit is associated with a spectrum range where the confinement
potential is well behaved (that is far from the logarithmic singularity $%
\lambda =0$), and where the density of states can be taken as being
approximately a constant on the scale of a few levels. In accordance with
this definition one has 
\begin{equation}
K_{\rm{bulk}}\left( s,s^{\prime }\right) =\lim_{s,s^{\prime }\rightarrow
\infty }K_{\rm{orig}}\left( s,s^{\prime }\right) ,  \label{pm.09}
\end{equation}
where $s$ and $s^{\prime }$ should, nevertheless, remain far enough from the
end point ${\cal D}_N$ of the spectrum support.

Taking this limit in Eq. (\ref{pm.09}), we arrive at the universal sine law 
\begin{equation}
K_{\rm{bulk}}\left( s,s^{\prime }\right) =\frac{\sin \left[ \pi \left(
s-s^{\prime }\right) \right] }{\pi \left( s-s^{\prime }\right) }
\label{skpm}
\end{equation}
deeply connected to the Wigner--Dyson level statistics \cite{BTW-1992}.

\subsubsection{Soft--edge scaling limit and the universal Airy law}

Soft--edge scaling limit is relevant to the tail of eigenvalue support where
crossover occurs from a nonzero density of states to a vanishing one. It is
known \cite{BB-1991,GM-1990} that by tuning coefficients $d_k$ which enter
the regular part $v$ of confinement potential (see Eq. (\ref{vr})), one can
obtain a macroscopic (Dyson) density of states which possesses a singularity
of the type $\nu _D\left( \lambda \right) \propto \left( 1-\lambda ^2/{\cal D%
}_N^2\right) ^{m+1/2}$ with the multicritical index $m=0,2,4,$ etc. (Odd
indices $m$ are inconsistent with our choice that the leading coefficient $%
d_p$, entering the regular component $v\left( \lambda \right) $ of
confinement potential, be positive in order to keep a convergence of
integral for partition function ${\cal Z}_N$ in Eq. (\ref{1s})). It was
shown in Ref. \cite{KF-1997a} within the Shohat method that as long as the
multicriticality of the order $m$ is reached, the eigenvalue correlations in
the vicinity of the soft edge become universal, and are independent of the
particular potential chosen. The order $m$ of the multicriticality is the
only parameter which governs spectral correlations in the soft--edge scaling
limit. Here, however, we restrict ourselves to a general confinement
potential without tuning to the multicritical point, that corresponds to $m=0
$. 

Let us move the spectrum origin to its end
point ${\cal D}_N$, making the replacement 
\begin{equation}
\lambda _s={\cal D}_N\left[ 1+\left( s/2\right) \left( 2(\pi {\cal D}_N{\cal %
R}_N(1))^{-1}\right) ^{2/3}\right] ,  \label{pm.12}
\end{equation}
that defines the {\it soft--edge scaling limit} provided $s\ll \left( {\cal D%
}_N{\cal R}_N\left( 1\right) \right) ^{2/3}\propto N^{2/3}$. It is
straightforward to show from Eqs. (\ref{pm.03}) and (\ref{pm.04}) that in
the vicinity of the end point ${\cal D}_N$ the function $\widehat{\varphi }%
_N\left( s\right) =\varphi _N^{\left( \alpha \right) }\left( \lambda _s-%
{\cal D}_N\right) $ obeys the universal differential equation 
\begin{equation}
\widehat{\varphi }_N^{\prime \prime }\left( s\right) -s\widehat{\varphi }%
_N\left( s\right) =0,  \label{pm.14}
\end{equation}
and that the following relation takes place, 
\begin{equation}
\widehat{\varphi }_{N-1}\left( s\right) =\widehat{\varphi }_N\left( s\right)
+\left( \frac 2{\pi {\cal D}_N{\cal R}_N\left( 1\right) }\right) ^{1/3}%
\widehat{\varphi }_N^{\prime }\left( s\right) .  \label{pm.15}
\end{equation}

Solution to Eq. (\ref{pm.14}) which decreases at $s\rightarrow +\infty $
(that is at far tails of the density of states) can be represented through
the Airy function 
\begin{equation}
\mathop{\rm Ai}
\left( s\right) =\frac 13\left\{ 
\begin{array}{cc}
s^{1/2}\left[ I_{-1/3}\left( \frac 23s^{3/2}\right) -I_{1/3}\left( \frac 23%
s^{3/2}\right) \right] , & s>0, \\ 
\left| s\right| ^{1/2}\left[ J_{-1/3}\left( \frac 23\left| s\right|
^{3/2}\right) +J_{1/3}\left( \frac 23\left| s\right| ^{3/2}\right) \right] ,
& s<0.
\end{array}
\right.  \label{g}
\end{equation}
as follows 
\begin{equation}
\widehat{\varphi }_N\left( s\right) =a{%
\mathop{\rm Ai}
}\left( s\right) ,  \label{pm.16}
\end{equation}
with $a$ being an unknown constant. Making use of Eq. (\ref{pm.15}), we
obtain that in the vicinity of the soft edge the scalar kernel is 
\begin{equation}
K_N\left( \lambda _s,\lambda _{s^{\prime }}\right) =b\frac{{%
\mathop{\rm Ai}
}\left( s\right) {%
\mathop{\rm Ai}
}^{\prime }\left( s^{\prime }\right) -{%
\mathop{\rm Ai}
}\left( s^{\prime }\right) {%
\mathop{\rm Ai}
}^{\prime }\left( s\right) }{s-s^{\prime }}{,}  \label{pm.17}
\end{equation}
where $b$ is an unknown constant again. It can be found by fitting \cite
{KF-1997} the density of states $K_N\left( \lambda _s,\lambda _s\right) $,
Eq. (\ref{pm.17}), to the Dyson density of states $\nu _D\left( \lambda
_s\right) $, Eq. (\ref{pm.025}), near the soft edge provided $1\ll s\ll
N^{2/3}$. This yields us the value $b=$ $c_N^{-1}(\pi c_N{\cal R}%
_N(1))^{2/3} $. Thus, we obtain that in the soft--edge scaling limit, Eq. (%
\ref{pm.12}), the scalar kernel $K_{\rm{{soft}}}\left( s,s^{\prime
}\right) =\lim_{N\rightarrow \infty }\lambda _s^{\prime }K_N\left( \lambda
_s,\lambda _{s^{\prime }}\right) $ satisfies the universal Airy law 
\begin{equation}
K_{\rm{{soft}}}\left( s,s^{\prime }\right) =\frac{{%
\mathop{\rm Ai}
}\left( s\right) {%
\mathop{\rm Ai}
}^{\prime }\left( s^{\prime }\right) -{%
\mathop{\rm Ai}
}\left( s^{\prime }\right) {%
\mathop{\rm Ai}
}^{\prime }\left( s\right) }{s-s^{\prime }}  \label{pm.18}
\end{equation}
which does not depend on the details of the confinement potential. In fact,
the Airy law is a particular case $\left( m=0\right) $ of more general
multicritical correlations characterized by the index $m$ of the
multicriticality. For more details we refer the reader to Refs. \cite
{KF-1997a,KF-1998}.

It follows from Eq. (\ref{pm.18}) that the density of states in the same
scaling limit 
\begin{equation}
\nu _{\rm{{soft}}}\left( s\right) =\left( \frac d{ds}{%
\mathop{\rm Ai}
}\left( s\right) \right) ^2-s\left[ {%
\mathop{\rm Ai}
}\left( s\right) \right] ^2  \label{pm.20}
\end{equation}
is also universal. The large--$\left| s\right| $ behavior of $\nu _{\rm{{%
soft}}}$ can be deduced from the known asymptotic expansions \cite{D-1973}
of the Bessel functions, 
\begin{equation}
\nu _{\rm{{soft}}}=\left\{ 
\begin{array}{ll}
\frac{\left| s\right| ^{1/2}}\pi -\frac 1{4\pi \left| s\right| }\cos \left( 
\frac 43\left| s\right| ^{3/2}\right) , & s\rightarrow -\infty , \\ 
\frac 1{8\pi s}\exp \left( -\frac 43s^{3/2}\right) , & s\rightarrow +\infty .
\end{array}
\right.  \label{pm.21}
\end{equation}
Note that the leading order behavior as $s\rightarrow -\infty $ is
consistent with the $\left| s\right| ^{1/2}$ singularity of the bulk density
of states. 

\subsection{Discussion}

Looking back at the formalism developed we should reiterate that the crucial
point in the derivations above is the large--$N$ limit, Eq. (\ref{a2}), of the
recurrence equation for associated orthogonal polynomials. It was precisely
this limit that led us to the important relation Eq. (\ref{pm.02}) and to
the effective Schr\"{o}dinger equation in the form of Eq. (\ref{pm.03})
which is a nonuniversal in general. However, it takes locally universal
forms in the spectrum bulk (where $\nu _D$ is approximatelly a constant on
the scale of a few eigenlevels), near the spectrum origin (where all the
nontrivial information is contained in $\lambda ^{-2}$ term in front of $%
\varphi _N^{\left( \alpha \right) }$ in Eq. (\ref{pm.03})), and near the
soft edge of the spectrum (where universality shows up in the universal
square--root singularity of $\nu _D$). These three locally universal
features of Eq. (\ref{pm.03}) have led us to the universal sine, Bessel and
Airy kernels in corresponding scaling limits.

We stress that Eq. (\ref{a2}) is, in fact, the leading--order--limit as $%
N\rightarrow \infty $. How accurate is this approximation, and do situations
exist where the next--order terms in the recurrence equation should be taken
into account? A partial answer to this question was given in Ref. \cite
{ADMN-1998} whose authors, remaining within the framework of the Shohat
method, convincingly demonstrated that corrections to Eq. (\ref{a2}) are of
importance in a problem of multicritical spectral correlations
near the spectrum origin. The effective Schr\"{o}dinger equation obtained
there for two particular matrix ensembles with fine--tuned confinement
potentials was shown not only to depend on the macroscopic spectral density $%
\nu _D\left( \lambda \right) $ but, in addition, to contain contributions
from subdominant terms in $1/N$ expansion for the recurrence coefficients.
It is important however, that in the situation in question the resulting
differential equation contained the universal functions ${\cal F}%
_N $ and ${\cal G}_N$ involving certain universal combinations of recurrence
coefficients and coupling constants responsible for the fine tuning of the
multicritical confinement potential. Having a complicated form, the
effective Schr\"{o}dinger equation could not be solved analytically, but,
remarkably, the authors of Ref. \cite{ADMN-1998} succeded in identifying a
certain ``mesoscopic'' limit, in which the numerical solution of the exact
differential equation and the analytical solution of the approximate
differential equation obtained by making use of the relation Eq. (\ref{pm.02}%
) were shown to have quite similar qualitative features. With increasing of
the order of the multicriticality near the spectrum origin, the approximate
(analytical) and exact (numerical) solutions were shown to approach each
other even quantitatively, demonstrating thus the potentialities of the
Shohat method even in its simplest formulation presented above.

\section{Two--Band Random Matrices}

\subsection{Multi--Band Spectral Regimes}

Ensembles of large random matrices ${\bf H}$ generated by the joint
distribution function $P\left[ {\bf H}\right] $, Eq. (\ref{mar.00}), may
display phase transitions under non--monotonic deformation of the
confinement potential $V\left[ {\bf H}\right] $. Different phases are
characterized by topologically different arrangements of eigenvalues in
random matrix spectra that may have multiple--band structure. Random
matrices, whose spectra undergo phase transitions, appear in quantizing
two--dimensional gravity \cite{M-1988,DSS-1990,SS-1991}, in the context of
quantum chromodynamics \cite{JNZ-1996,JV-1996}, as well as in some models of
particles interacting in high dimensions \cite{CKPR-1995}. Transition
regimes realized in invariant random matrix ensembles have implications for
a certain class of Calogero--Sutherland--Moser models \cite{MHK-1995}. These
matrix models may also be applicable to chaotic systems having a forbidden
gap in the energy spectrum.

It is convenient to parametrize the confinement potential $V\left( \lambda
\right) $ entering Eq. (\ref{mar.00}) by a set of coupling constants $%
\left\{ d\right\} =\left\{ d_1,...,d_p\right\} $, 
\begin{equation}
V\left( \lambda \right) =\sum_{k=1}^p\frac{d_k}{2k}\lambda ^{2k},\;d_p>0,
\label{tbq.02}
\end{equation}
so that we may consider the phase transitions as occurring in $\left\{
d\right\} $--space. Because the confinement potential is an even function,
the associated random matrix model possesses so--called $Z2$--symmetry.

Variations of the coupling constants affect the Dyson density $\nu _D$, that
can be found by minimizing the free energy $F_N=-\log {\cal Z}_N$, Eq. (\ref
{mar.00}), subject to a normalization constraint $\int \nu _D\left( \lambda
\right) d\lambda =N$, 
\begin{equation}
\frac{dV}{d\lambda }-{\cal P}\int dt\frac{\nu _D\left( t\right) }{\lambda -t}%
=0,  \label{tbq.03}
\end{equation}
where ${\cal P}$ indicates a principal value of the integral. When all $d_k$
are positive, so that confinement potential is monotonic, the spectral
density $\nu _D$ has a single--band support, ${\cal N}_{\rm{b}}=1$.
Non--monotonic deformation of the confinement potential can be carried out
by changing the signs of some of $d_k$ $\left( k\neq p\right) $. Such a {\it %
continuous} variation of coupling constants may lead, under certain
conditions, to a {\it discontinuous} change of the topological structure of
spectral density $\nu _D$, when the eigenvalues $\left\{ \lambda \right\} $
are arranged in ${\cal N}_{\rm{b}}>1$ ``allowed'' bands separated by
``forbidden'' gaps.

The phase structure of Hermitean $\left( \beta =2\right) $ one--matrix model
Eq. (\ref{mar.00}) has been studied in a number of works \cite
{S-1982,CMM-1990,J-1991,DDJT-1990}, where the simplest examples of
non--monotonic quartic and sextic confinement potentials have been examined.
It has been found that there are domains in the phase space of coupling
constants where only a particular solution for $\nu _D$ exists, and it has a
fixed number ${\cal N}_{\rm{b}}$ of allowed bands. However, in some
regions of the phase space, one can have more than one kind of solution of
the saddle--point equation Eq. (\ref{tbq.03}). In this situation, solutions
with different number of bands ${\cal N}_{\rm{b}}^{\left( 1\right) },$ $%
{\cal N}_{\rm{b}}^{\left( 2\right) },\ldots $ are present simultaneously.
When such an overlap appears, one of the solutions, say ${\cal N}_{\rm{b}%
}^{\left( k\right) }$, has the lowest free energy $F_N^{\left( k\right) }$,
and this solution is dominant, while the others are subdominant. Moreover,
numerical calculations \cite{J-1991} showed that some special regimes exist
in which the {\it bulk} spectral density obtained as a solution to the
saddle--point equation Eq. (\ref{tbq.03}) differs significantly from the
genuine level density computed numerically within the framework of the
orthogonal polynomial technique. It was then argued that such a genuine
density of levels cannot be interpreted as a multi--band solution with an
integer number of bands. A full understanding of this phenomenon is still
absent.

Recently, interest was renewed in multi--band regimes in invariant random
matrix ensembles. An analysis based on a loop equation technique \cite
{AA-1996,A-1996} showed that fingerprints of phase transitions appear not
only in the Dyson density but also in the (universal) wide--range eigenvalue
correlators, which in the multi--band phases differ from those known in the
single--band phase \cite{AJM-1990,BZ-1993,B-1994}. A renormalization group
approach developed in Ref. \cite{HINS-1997} supported the results found in
Refs. \cite{AA-1996,A-1996} for the particular case of two allowed bands,
referring a new type of universal wide--range eigenlevel correlators to an
additional attractive fixed point of a renormalization group transformation.

As it was already stressed in the Introduction, the method of loop equations 
\cite{AA-1996,A-1996}, used for a treatment of non--Gaussian, unitary
invariant, random matrix ensembles fallen in a multi--band phase, is only
suitable for computing the global characteristics of spectrum. Therefore, an
appropriate approach is needed capable of analyzing local characteristics of
spectrum (manifested on the scale of a few eigenlevels). A possibility to
probe the local properties of eigenspectrum is offered by the method of
orthogonal polynomials. A step in this direction was taken in the paper 
\cite{D-1997}, where an ansatz was proposed for large--$N$ asymptotes of
orthogonal polynomials associated with a random matrix ensemble having two
allowed bands in its spectrum. Because the asymptotic formula proposed there
is of the Plancherel--Rotach type \cite{S-1967}, it is only applicable for
studying eigenvalue correlations in the spectrum bulk and cannot be used for
studying local correlations in an arbitrary spectrum range (for example,
near the edges of two--band eigenvalue support).

Below we demonstrate that the Shohat method needs minimal modifications to
allow a unified treatment of eigenlevel correlations in the unitary
invariant $%
\mathop{\rm U}
\left( N\right) $ matrix model $\left( \beta =2\right) $ with a forbidden
gap. In particular, we will be able to study both the fine structure of
local characteristics of the spectrum in different scaling limits and
smoothed global spectral correlations. As is the single--band phase, the
treatment presented below is based on the direct reconstruction of spectral
correlations from the recurrence equation for the corresponding orthogonal
polynomials.

\subsection{Effective Schr\"{o}dinger Equation in the Two--Band Phase and
Local Eigenvalue Correlations}

Let us consider the situation when the confinement potential has two deep
wells leading to the Dyson density supported on two disjoint intervals
located symmetrically about the origin, ${\cal D}_N^{-}<\left| \lambda
\right| <{\cal D}_N^{+}$. In this situation, the recurrence coefficients $%
c_n $ entering Eq. (\ref{eq.16}) are known to be double--valued functions of
the number $n$ \cite{M-1988,DDJT-1990}. This means that for $n=N\gg 1$ and
in contrast with a single--band phase, one must distinguish between
coefficients $c_{N\pm 2q}\approx c_N$ and coefficients $c_{N-1\pm 2q}\approx
c_{N-1}$, belonging to two different, smooth (in index) sub--sequences;
here, integer $q\sim {\cal O}\left( N^0\right) $. Bearing this in mind, the
large--$N$ version of recurrence equation Eq. (\ref{eq.16}) can be rewritten
as 
\begin{equation}
\left[ \lambda ^2-\left( c_N^2+c_{N-1}^2\right) \right] P_N\left( \lambda
\right) =c_Nc_{N-1}\left[ P_{N-1}\left( \lambda \right) +P_{N+1}\left(
\lambda \right) \right] ,  \label{tbq.19}
\end{equation}
whence the two analogues of the asymptotic expansion Eq. (\ref{a3}) can be
obtained. They are given by Eqs. (\ref{b2}) and (\ref{b3}) of Appendix B. We notice that
this is the only point crucial for extending the Shohat method to the
double--well matrix models considered in this Section. These new expansions
make it possible to compute the required functions ${\cal F}_N$ and ${\cal G}%
_N$ entering the differential equation Eq. (\ref{eq.30}) for fictitious
wave functions in the limit $N\gg 1$.

In accordance with the general framework of the Shohat method, we have to
compute two functions (compare with Eqs. (\ref{eq.A}) and (\ref{eq.B})) 
\begin{eqnarray}
A_N\left( \lambda \right) &=&2c_N\sum_{k=1}^pd_k\sum_{\sigma =1}^k\Lambda
_{2\left( k-\sigma \right) }^{\left( N\right) }\lambda ^{2\sigma -2},
\label{tbeq.A} \\
B_N\left( \lambda \right) &=&2c_N\sum_{k=1}^pd_k\sum_{\sigma =1}^k\Gamma
_{2\left( k-\sigma \right) -1}^{\left( N\right) }\lambda ^{2\sigma -1},
\label{tbeq.B}
\end{eqnarray}
involving the objects $\Lambda _{2\sigma }^{\left( N\right) }$ and $\Gamma
_{2\sigma +1}^{\left( N\right) }$, for which there exist the useful integral
representations given by Eqs. (\ref{b7}) and (\ref{b15}) in Appendix B. Substituting
them into Eqs. (\ref{tbeq.A}) and (\ref{tbeq.B}) one is able to perform the
double summation over indices $k$ and $\sigma $. Omiting details of
straightforward calculations we present the final answer given by the
formulas 
\begin{eqnarray}
A_N\left( \lambda \right)  &=&\frac 2\pi \left( {\cal D}_N^{+}-\left(
-1\right) ^N{\cal D}_N^{-}\right) {\cal P}\int_{{\cal D}_N^{-}}^{{\cal D}%
_N^{+}}\frac{dV}{dt}\frac{t^2}{t^2-\lambda ^2}  \nonumber  \label{tbq.24} \\
&&\times \frac{dt}{\left[ \left( {\cal D}_N^{+}\right) ^2-t^2\right]
^{1/2}\left[ t^2-\left( {\cal D}_N^{-}\right) ^2\right] ^{1/2}},
\label{tbq.24}
\end{eqnarray}
\begin{equation}
B_N\left( \lambda \right) =\frac{2\lambda }\pi {\cal P}\int_{{\cal D}%
_N^{-}}^{{\cal D}_N^{+}}\frac{dV}{dt}\frac{t^2-\left( -1\right) ^N{\cal D}%
_N^{-}{\cal D}_N^{+}}{\left[ \left( {\cal D}_N^{+}\right) ^2-t^2\right]
^{1/2}\left[ t^2-\left( {\cal D}_N^{-}\right) ^2\right] ^{1/2}}\frac{dt}{%
t^2-\lambda ^2}-\frac{dV}{d\lambda }.  \label{tbq.25}
\end{equation}
Note, that along with a different (compared to the single--band phase)
functional form of the functions $A_N\left( \lambda \right) $ and $B_N\left(
\lambda \right) $, these functions are,
in fact, double--valued in index $N$, and behave in a different way for odd
and even $N$. This is a direct consequence of the ``period--two'' behavior 
\cite{M-1988,DDJT-1990} of the recurrence coefficients $c_n$.

Having obtained the explicit expressions for functions $A_N$ and $B_N$, it
is easy to verify that coefficients ${\cal F}_n\left( \lambda \right) $ and $%
{\cal G}_n\left( \lambda \right) $ entering the differential equation Eq. (%
\ref{eq.30}) for the fictitious wave function $\varphi _n\left( \lambda
\right) $ may be expressed in terms of the Dyson density $\nu _D^{\left( 
\rm{II}\right) }$ in the two--cut phase supported on two disconnected
intervals $\lambda \in \left( -{\cal D}_N^{+},-{\cal D}_N^{-}\right) \cup
\left( {\cal D}_N^{-},{\cal D}_N^{+}\right) $, 
\begin{equation}
\nu _D^{\left( \rm{II}\right) }\left( \lambda \right) =\frac 2{\pi ^2}%
\left| \lambda \right| {\cal P}\int_{{\cal D}_N^{-}}^{{\cal D}_N^{+}}dt\frac{%
dV/dt}{t^2-\lambda ^2}\left( \frac{\left( {\cal D}_N^{+}\right) ^2-\lambda ^2%
}{\left( {\cal D}_N^{+}\right) ^2-t^2}\right) ^{1/2}\left( \frac{\lambda
^2-\left( {\cal D}_N^{-}\right) ^2}{t^2-\left( {\cal D}_N^{-}\right) ^2}%
\right) ^{1/2}  \label{tbq.26}
\end{equation}
when $N\gg 1$. This formula can be obtained either via the procedure of the
Sec. 4.1 or within the mean--field approach, Eq. (\ref{tbq.03}). Here $%
{\cal D}_N^{-}$ and ${\cal D}_N^{+}$ are the end points of the eigenvalue
support that obey the two integral equations 
\begin{equation}
\int_{{\cal D}_N^{-}}^{{\cal D}_N^{+}}\frac{dV}{dt}\frac{t^2dt}{\left[
\left( {\cal D}_N^{+}\right) ^2-t^2\right] ^{1/2}\left[ t^2-\left( {\cal D}%
_N^{-}\right) ^2\right] ^{1/2}}=\frac{\pi N}2,  \label{tbee.23}
\end{equation}
\begin{equation}
\int_{{\cal D}_N^{-}}^{{\cal D}_N^{+}}\frac{dV}{dt}\frac{dt}{\left[ \left( 
{\cal D}_N^{+}\right) ^2-t^2\right] ^{1/2}\left[ t^2-\left( {\cal D}%
_N^{-}\right) ^2\right] ^{1/2}}=0,  \label{tbee.24}
\end{equation}
derived in Appendix C.

By making use of the Eqs. (\ref{eq.31a}), (\ref{eq.31b}) and (\ref{tbq.24})
-- (\ref{tbq.26}) we obtain in the large--$N$ limit 
\begin{equation}
{\cal F}_N\left( \lambda \right) =\frac d{d\lambda }\log \left( \frac{\pi
\left| \lambda \right| \nu _D^{\left( \rm{II}\right) }\left( \lambda
\right) }{\left[ \left( {\cal D}_N^{+}\right) ^2-\lambda ^2\right]
^{1/2}\left[ \lambda ^2-\left( {\cal D}_N^{-}\right) ^2\right] ^{1/2}}%
\right) ,  \label{tbq.27}
\end{equation}
\begin{equation}
{\cal G}_N\left( \lambda \right) =\left( \pi \nu _D^{\left( \rm{II}\right)
}\left( \lambda \right) \right) ^2,  \label{tbq.28}
\end{equation}
so that for $N\gg 1$ the effective Schr\"{o}dinger equation in the two--cut
phase reads \cite{KF-1998a}  
\begin{eqnarray}
\frac{d^2\varphi _N\left( \lambda \right) }{d\lambda ^2} &-&\left[ \frac d{%
d\lambda }\log \left( \frac{\pi \left| \lambda \right| \nu _D^{\left( \rm{%
II}\right) }\left( \lambda \right) }{\left[ \left( {\cal D}_N^{+}\right)
^2-\lambda ^2\right] ^{1/2}\left[ \lambda ^2-\left( {\cal D}_N^{-}\right)
^2\right] ^{1/2}}\right) \right] \frac{d\varphi _N\left( \lambda \right) }{%
d\lambda }  \nonumber  \label{tbq.29} \\
&+&\left( \pi \nu _D^{\left( \rm{II}\right) }\left( \lambda \right)
\right) ^2\varphi _N\left( \lambda \right) =0.  \label{tbq.29}
\end{eqnarray}
As ${\cal D}_N^{-}$ tends to zero, we reproduce the equation (\ref{pm.03})
with $\alpha =0$ valid in the single--band regime.

Local eigenvalue correlations in the spectra of two--band random matrices are
completely determined by the Dyson density of states entering the effective
Schr\"{o}dinger equation Eq. (\ref{tbq.29}).

(i) In the spectrum bulk, the Dyson density is a well behaved function that
can be taken approximately as being a constant on the scale of a few
eigenlevels. Then, in the vicinity of some $\lambda _0$ that is chosen to be
far enough from the spectrum end points $\pm {\cal D}_N^{\pm }$, Eq. (\ref
{tbq.29}) takes the form 
\begin{equation}
\frac{d^2\varphi _N\left( \lambda \right) }{d\lambda ^2}+\left[ \pi /\Delta
\left( \lambda _0\right) \right] ^2\varphi _N\left( \lambda \right) =0,
\label{sin.1}
\end{equation}
with $\Delta \left( \lambda _0\right) =1/\nu _D^{\left( \rm{II}\right)
}\left( \lambda _0\right) $ being the mean level spacing in the vicinity of $%
\lambda _0$. Clearly, the universal sine law, Eq. (\ref{skpm}), for the
two--point kernel follows immediately.

(ii) To study the eigenvalue correlations near the end points of an
eigenvalue support we notice that in the absence of the fine tunning of
confinement potential, the Dyson density has a universal square--root
singularity in the vicinity of $\left| \lambda \right| ={\cal D}_N^{\pm }$, that is
$\nu _D^{\left( \rm{II}\right) }\left( \lambda \right) \propto
\left( 1-\left( \lambda /{\cal D}_N^{\pm }\right) ^2\right) ^{1/2}$. We then readily recover the
universal Airy correlations, Eq. (\ref{pm.18}), previously found in the
soft--edge scaling limit for $%
\mathop{\rm U}
\left( N\right) $ invariant matrix model in the single--band phase. 

\subsection{Global Connected ``Density--Density'' Correlator}

Let us turn to the study of the smoothed connected ``density--density'' correlator
that is expressed in terms of the scalar kernel as follows (see Eq. (\ref
{wrmequ.13})), 
\begin{eqnarray}
\rho _{c\rm{II}}^{\left( N\right) }\left( \lambda ,\lambda ^{\prime
}\right)  &=&-\frac{c_N^2}{\left( \lambda -\lambda ^{\prime }\right) ^2}%
\left\{ \overline{\varphi _N^2\left( \lambda \right) }\,\overline{\varphi
_{N-1}^2\left( \lambda ^{\prime }\right) }+\overline{\varphi _N^2\left(
\lambda ^{\prime }\right) }\,\overline{\varphi _{N-1}^2\left( \lambda
\right) }\right.   \nonumber \\
&&\left. -2\overline{\varphi _N\left( \lambda \right) \varphi _{N-1}\left(
\lambda \right) }\,\overline{\varphi _N\left( \lambda ^{\prime }\right)
\varphi _{N-1}\left( \lambda ^{\prime }\right) }\right\} .  \label{tbee.33}
\end{eqnarray}
Here $\lambda \neq \lambda ^{\prime }$. Equation (\ref{tbee.33}) contains
(before averaging) rapid oscillations on the scale of the mean level spacing. These
oscillations are due to presence of oscillating wave functions $\varphi _N$
and $\varphi _{N-1}$.

To average over the rapid oscillations, we integrate, over the entire real
axis, rapidly varying wave functions in Eq. (\ref{tbee.33}) multiplied by an
arbitrary, smooth, slowly varying function, which without any loss of
generality can be choosen to be $\lambda ^{2s}$ for $\varphi _N^2\left(
\lambda \right) $ and $\lambda ^{2s+1}$ for $\varphi _N\left( \lambda
\right) \varphi _{N-1}\left( \lambda \right) $ ($s$ is an arbitrary positive
integer, $s>0$). Consider, first, the integral 
\begin{equation}
I_1^{\left( N\right) }=\int_{-\infty }^{+\infty }d\lambda \lambda
^{2s}\varphi _N^2\left( \lambda \right) =\Lambda _{2s}^{\left( N\right) }.
\label{tbee.34}
\end{equation}
With the help of Eq. (\ref{b7}), and bearing in mind that $\varphi _N^2\left(
\lambda \right) $ is an even function, we conclude that 
\begin{equation}
\ I_1^{\left( N\right) }=\frac 1\pi \int_{{\cal D}_N^{-}<\left| \lambda
\right| <{\cal D}_N^{+}}\frac{\left| \lambda \right| \lambda ^{2s}d\lambda }{%
\left[ \left( {\cal D}_N^{+}\right) ^2-\lambda ^2\right] ^{1/2}\left[
\lambda ^2-\left( {\cal D}_N^{-}\right) ^2\right] ^{1/2}},  \label{tbee.38}
\end{equation}
whence, in the large--$N$ limit, 
\begin{equation}
\overline{\varphi _N^2\left( \lambda \right) }=\frac{\Omega _\lambda }\pi 
\frac{\left| \lambda \right| }{\left[ \left( {\cal D}_N^{+}\right)
^2-\lambda ^2\right] ^{1/2}\left[ \lambda ^2-\left( {\cal D}_N^{-}\right)
^2\right] ^{1/2}}.  \label{tbee.39}
\end{equation}
Here
\begin{equation}
\Omega _\lambda =\Theta \left( {\cal D}_N^{+}-\left| \lambda \right| \right)
\Theta \left( \left| \lambda \right| -{\cal D}_N^{-}\right)   \label{omega}
\end{equation}
with $\Theta $ being a step function. The same procedure should be carried
out with the expression $\varphi _N\left( \lambda \right) \varphi
_{N-1}\left( \lambda \right) $ in Eq. (\ref{tbee.33}). Since this
construction is an odd function of $\lambda $, we have to consider the
integral 
\begin{equation}
I_2^{\left( N\right) }=\int_{-\infty }^{+\infty }d\lambda \lambda
^{2s+1}\varphi _N\left( \lambda \right) \varphi _{N-1}\left( \lambda \right)
=\Gamma _{2s+1}^{\left( N\right) }.  \label{tbee.40}
\end{equation}
With the help of Eq. (\ref{b15}), and exploiting the oddness of $\varphi _N\left(
\lambda \right) \varphi _{N-1}\left( \lambda \right) $, we rewrite Eq. (\ref
{tbee.40}) in the form 
\begin{eqnarray}
I_2^{\left( N\right) } &=&\frac 1{\pi \left[ {\cal D}_N^{+}-\left( -1\right)
^N{\cal D}_N^{-}\right] }  \nonumber  \label{tbee.44} \\
&&\times \int_{{\cal D}_N^{-}<\left| \lambda \right| <{\cal D}_N^{+}}\frac{%
\left[ \lambda ^2-\left( -1\right) ^N{\cal D}_N^{-}{\cal D}_N^{+}\right] 
\mathop{\rm sgn}
\left( \lambda \right) d\lambda }{\left[ \left( {\cal D}_N^{+}\right)
^2-\lambda ^2\right] ^{1/2}\left[ \lambda ^2-\left( {\cal D}_N^{-}\right)
^2\right] ^{1/2}},  \label{tbee.44}
\end{eqnarray}
whence 
\begin{eqnarray}
\overline{\varphi _N\left( \lambda \right) \varphi _{N-1}\left( \lambda
\right) } &=&\frac{\Omega _\lambda 
\mathop{\rm sgn}
\left( \lambda \right) }{\pi \left[ {\cal D}_N^{+}-\left( -1\right) ^N{\cal D%
}_N^{-}\right] }  \nonumber  \label{tbee.45} \\
&&\times \frac{\lambda ^2-\left( -1\right) ^N{\cal D}_N^{-}{\cal D}_N^{+}}{%
\left[ \left( {\cal D}_N^{+}\right) ^2-\lambda ^2\right] ^{1/2}\left[
\lambda ^2-\left( {\cal D}_N^{-}\right) ^2\right] ^{1/2}}.  \label{tbee.45}
\end{eqnarray}
Combining Eqs. (\ref{tbee.33}), (\ref{tbee.39}), (\ref{tbee.45}) and (\ref{b14}),
we finally arrive at the following formula for smoothed ``density--density''
correlator \cite{KF-1998a} 
\begin{eqnarray}
\rho _{c\rm{II}}^{\left( N\right) }\left( \lambda ,\lambda ^{\prime
}\right)  &=&-\frac{%
\mathop{\rm sgn}
\left( \lambda \lambda ^{\prime }\right) }{2\pi ^2}\frac{\Omega _\lambda
\Omega _{\lambda ^{\prime }}}{Q_N\left( \lambda \right) Q_N\left( \lambda
^{\prime }\right) }\left\{ \left( -1\right) ^N{\cal D}_N^{-}{\cal D}%
_N^{+}\right.   \label{y1} \\
&&\left. +\frac 1{\left( \lambda -\lambda ^{\prime }\right) ^2}\left[
\lambda \lambda ^{\prime }-({\cal D}_N^{-})^2\right] \left[ ({\cal D}%
_N^{+})^2-\lambda \lambda ^{\prime }\right] \right\} ,  \nonumber  \label{y1}
\end{eqnarray}
where
\begin{equation}
Q_N\left( \lambda \right) =\left[ ({\cal D}_N^{+})^2-\lambda ^2\right]
^{1/2}\left[ \lambda ^2-({\cal D}_N^{-})^2\right] ^{1/2}.  \label{y101}
\end{equation}

It is seen from Eq. (\ref{y1}) that smoothed ``density--density'' correlator
in the two--band phase differs from that in the single--band phase, Eq. (\ref
{sc.04}). However, it is still universal in the sense that the information
of the distribution Eq. (\ref{mar.00}) is encoded into the
``density--density'' correlator only through the end points ${\cal D}_N^{\pm
}$ of the eigenvalue support. The striking parity effect in the new
universal function Eq. (\ref{y1}), that is the {\it sharp} dependence of
correlations on the oddness/evenness of the dimension $N$ of the random
matrices, is the main {\it qualitative} difference as compared to the global
corelations in random matrices fallen in the single--band phase. This effect
is most pronounced in the case of unbounded spectrum. The origin of this
unusual large--$N$ behavior will be discussed later on.

Finally, let us speculate about the universal correlator Eq. (\ref{y1}) in
the limit of {\it unbounded} spectrum, ${\cal D}_N^{+}\rightarrow \infty $,
with a gap. Inasmuch as it describes correlations between the eigenlevels
which are repelled from each other in accordance with the logarithmic law,
that is known to be realized \cite{AS-1986,JPB-1993} in weakly disordered
systems on the energy scale $\left| \lambda -\lambda ^{\prime }\right| \ll
E_c$ ($E_c$ is the Thouless energy), we may conjecture that the
corresponding limiting universal expression 
\begin{eqnarray}
\lim_{{\cal D}_N^{+}\rightarrow +\infty }\rho _{c\rm{II}}^{\left( N\right)
}\left( \lambda ,\lambda ^{\prime }\right)  &=&-\frac{%
\mathop{\rm sgn}
\left( \lambda \lambda ^{\prime }\right) }{2\pi ^2\left( \lambda -\lambda
^{\prime }\right) ^2}\Theta \left( \left| \lambda \right| -\Delta \right)
\Theta \left( \left| \lambda ^{\prime }\right| -\Delta \right)   \nonumber
\label{ee.48} \\
&&\times \frac{\lambda \lambda ^{\prime }-\Delta ^2}{\left[ \lambda
^2-\Delta ^2\right] ^{1/2}\left[ \lambda ^{\prime 2}-\Delta ^2\right] ^{1/2}}%
,  \label{ee.48}
\end{eqnarray}
reflects the universal properties of real chaotic systems with a forbidden
gap $\Delta ={\cal D}_N^{-}$ and broken time reversal symmetry, provided $%
\left| \lambda -\lambda ^{\prime }\right| \ll E_c$. In two limiting
situations (i) of gapless spectrum, $\Delta =0$, and (ii) far from the gap, $%
\left| \lambda \right| ,\left| \lambda ^{\prime }\right| \gg \Delta $, the
correlator Eq. (\ref{ee.48}) coincides with that known in the Random Matrix
Theory of gapless ensembles \cite{BZ-1993,B-1994} and derived in Ref. \cite
{AS-1986} within the framework of diagrammatic technique for spectrum of
electron in a random impurity potential.

\subsection{Discussion}

In this Section we have demonstrated how the Shohat method should be
transformed in order to study both global and local spectral characteristics
of $%
\mathop{\rm U}
\left( N\right) $ invariant ensembles of large random matrices possessing $%
Z2 $--symmetry, and deformed in such a way that their spectra contain a
forbidden gap. We proved that in the pure two--band phase, the local
eigenvalue correlations are insensitive to this deformation both in the bulk
and soft--edge scaling limits. In contrast, global smoothed eigenvalue
correlations in the two--band phase differ drastically from those in the
single--band phase, and generically satisfy a universal law, Eq. (\ref{y1}),
which is unusually sensitive to the oddness/evenness of the random matrix
dimension provided the spectrum support is bounded. On the formal level,
this sensitivity is a direct consequence of the ``period--two'' behavior 
\cite{M-1988,DDJT-1990} of the recurrence coefficients $c_n$ that is
characteristic of two--band phase of reduced Hermitean matrix model. To see
this, consider the simplest connected correlator $\left\langle 
\mathop{\rm Tr}
{\bf H}%
\mathop{\rm Tr}
{\bf H}\right\rangle _c$ that can be {\it exactly} represented in terms of
recurrence coefficients for any $n$, 
\begin{equation}
\left\langle 
\mathop{\rm Tr}
{\bf H}%
\mathop{\rm Tr}
{\bf H}\right\rangle _c=c_n^2\rm{.}  \label{tbe.100}
\end{equation}
Since in the two--band phase $c_n$ is a double--valued function of index $n$%
, alternating between two different functions as $n$ goes from odd to even,
the large--$N$ limit of the correlator $\left\langle 
\mathop{\rm Tr}
{\bf H}%
\mathop{\rm Tr}
{\bf H}\right\rangle _c$ strongly depends on whether infinity is approached
through odd or even $N$. Then, an implementation of a double--valued
behavior of $c_n$ into the higher order correlators of the form $%
\left\langle 
\mathop{\rm Tr}
{\bf H}^k%
\mathop{\rm Tr}
{\bf H}^l\right\rangle _c$ contributing to the connected
``density--density'' correlator gives rise to the universal expression Eq. (%
\ref{y1}), which is valid for the two--band random matrix model with pure
Z2--symmetry.

Let us, however, point out that no such sensitivity has been detected in a
number of previous studies \cite{AA-1996,A-1996} exploiting a loop equation
technique. One possible explanation comes from the following reasons. In the
method of loop equations, used for a treatment of non--Gaussian random
matrix ensembles fallen in a multi--band phase, one has to keep the most
general (non--symmetric) confinement potential $V\left( \lambda \right)
=\sum_{k=1}^{2p}\widetilde{d}_k\lambda ^k/k$ until very end of the
calculations, and to take the thermodynamic limit $N\rightarrow \infty $
prior to any others. Therefore, $Z2$--symmetry in this calculational scheme
can only be implemented by restoring $Z2$--symmetry at the final stage of
the calculations, setting all the extra coupling constants $\widetilde{d}%
_{2k+1}$ to zero. Doing so, one arrives at the results reported in Refs. 
\cite{AA-1996,A-1996}. From this point of view, our treatment corresponds to
the {\it opposite sequence }of thermodynamic and $Z2$--symmetry limits,
since we have considered the random matrix model that possesses $Z2$%
--symmetry from the beginning. Qualitatively different large--$N$ behavior
of the smoothed connected ``density--density'' correlator, Eq. (\ref{y1}),
and of the smoothed connected two--point Green's function given by Eq. (15)
of Ref. \cite{A-1996} provides a direct evidence that the order of
thermodynamic and $Z2$--symmetry limits is indeed important when studying
global spectral characteristics of multi--band random matrices.

The parity effect manifested in global spectral correlators of double--well
matrix models was the focus of the discussion in the recent study \cite
{BD-1998}. The authors of Ref. \cite{BD-1998} noted that, contrary to the method of orthogonal polynomials, the standard large--%
$N$ limit techniques of analyzing matrix models like the loop equation
method \cite{AJM-1990,AA-1996} and the renormalization group approach \cite{BZJ-1992}
assume a smooth behavior with respect to $N$ in the thermodynamic limit. The
result Eq. (\ref{y1}) obtained by the authors of Ref. \cite{BD-1998} in a
different way \cite{D-1997} led them to conclusion that these methods need
to be revisited when one deals with matrix models posessing eigenvalue gaps.

\section{Conclusion}

In this review we presented a formalism for statistical description of
spectra of $%
\mathop{\rm U}
\left( N\right) $ invariant ensembles of large random matrices. It lies
within the general framework of the orthogonal polynomials' technique, and
consists of the direct reconstruction of spectral densities and spectral
correlations from the recurrence equation for orthogonal polynomials
associated with a given random matrix ensemble. We have demonstrated the
potentialities of this method, considering in a unified way both global and
local spectral characteristics in matrix models with and without an
eigenvalue gap. Although we directed our main attention to the most known bulk,
origin and soft--edge scaling limits characterized by the universal sine,
Bessel and Airy kernels, respectively, there are examples in the recent
literature signaling about applicability of the described formalism to more
refined situations -- such as multicritical correlations near the soft edge
of the spectrum support \cite{KF-1997a} and near the spectrum origin \cite
{ADMN-1998}.

Attaching special significance to the study of the large--$N$ limit of the
recurrence equation for associated orthogonal polynomials, this method turns
the recurrence equation into a kind of laboratory allowing the construction
of matrix models with nonstandard properties -- for example with eigenvalue
gaps -- by guessing a particular ansatz for the behavior of the recurrence
coefficients $c_N$ in the thermodynamic limit. Just this feature of the
formalism presented forces us, finally, to mention a crucial difference
between the random matrix ensembles with strong level confinement considered
here and the random matrix ensembles with extremely soft level confinement.
While the former ensembles (with confinement potentials of the Freud and
Erd\"{o}s type \cite{FKY-1996a}) are characterized by a powerlike large--$N$
limit of recurrence coefficients, $c_N\propto N^\rho $ $\left( \rho
>0\right) $, the latter (representing a class of $q$--deformed potentials 
\cite{MCIN-1993,BBP-1997}) exhibit a qualitatively different, exponential
rate of growth, $c_N\propto q^N$ $\left( q>1\right) $. This results in a
different large--$N$ limit of the recurrence equation that will not already
be as simple as stated in Eq. (\ref{a2}), and the emergence of different
nontrivial classes of spectral statistics is inevitable. We consider a
treatment of $q$--deformed random matrix ensembles as a challenge to the
Shohat method which, going back to 1930, had to wait so long to find its
application in Random Matrix Theory.
\vspace{0.5cm}
\begin{center}
{\bf Acknowledgments}
\end{center}
One of the authors (E. K.) thanks G. Akemann for useful discussions at the
Abdus Salam International Centre for Theoretical Physics and for
collaboration. Financial support through the Rothschild Postdoctoral
Fellowship (E. K.) is gratefully acknowledged.

\appendix 
\section{Integral Representation of $\Lambda _{2\sigma }^{\left( N\right) }$
and $\Gamma _{2\sigma +1}^{\left( N\right) }$: Single--Band Phase}

Consider the integral 
\begin{equation}
\Lambda _{2\sigma }^{\left( N\right) }=\int d\alpha \left( t\right) \left(
P_N^{\left( \alpha \right) }\left( t\right) \right) ^2t^{2\sigma }
\label{a1}
\end{equation}
with integer $\sigma \geq 0$. In the large--$N$ limit an alternative
explicit integral representation can be found for $\Lambda _{2\sigma
}^{\left( N\right) }$. This is achieved by making use of the large--$N$
version of the recurrence equation (\ref{eq.16}). It is known that in the
single--band phase of the matrix model the recurrence coefficients approach 
\cite{M-1988,DDJT-1990} a smooth (in index $N$) single--valued function, so
that $c_{N+q}\approx c_N$ for $q$ being of order ${\cal O}\left( N^0\right) $%
. Within this approximation one obtains from Eq. (\ref{eq.16}) 
\begin{equation}
\lambda P_N^{\left( \alpha \right) }\left( \lambda \right) =c_N\left(
P_{N-1}^{\left( \alpha \right) }\left( \lambda \right) +P_{N+1}^{\left(
\alpha \right) }\left( \lambda \right) \right) ,  \label{a2}
\end{equation}
whence it follows that 
\begin{equation}
\lambda ^mP_N^{\left( \alpha \right) }\left( \lambda \right)
=c_N^m\sum_{j=0}^m\left( 
\begin{array}{c}
m \\ 
j
\end{array}
\right) P_{N+2j-m}^{\left( \alpha \right) }\left( \lambda \right) ,\ m\geq 1.
\label{a3}
\end{equation}
The identity (\ref{a3}) can be proven by the mathematical induction. The advantage
of the asymptotic expansion Eq. (\ref{a3}) is that being substituted (for $%
m=2\sigma $) into Eq. (\ref{a1}), it immediately allows us to explicitly perform
the integration due to the orthogonality property Eq. (\ref{req.07}). This
yields 
\begin{equation}
\Lambda _{2\sigma }^{\left( N\right) }=c_N^{2\sigma }\sum_{j=0}^{2\sigma
}\left( 
\begin{array}{c}
2\sigma  \\ 
j
\end{array}
\right) \delta _{2j}^{2\sigma },  \label{a4}
\end{equation}
with $\delta _j^k$ being the Kronecker symbol. To evaluate the sum in Eq.
(\ref{a4}) we make use of the integral representation of the Kronecker symbol, 
\begin{equation}
\delta _j^k=%
\mathop{\rm Re}
\int_0^{2\pi }\frac{d\theta }{2\pi }\exp \left\{ i\theta \left( j-k\right)
\right\} .  \label{a5}
\end{equation}
We then find that 
\begin{equation}
\Lambda _{2\sigma }^{\left( N\right) }=c_N^{2\sigma }%
\mathop{\rm Re}
\int_0^{2\pi }\frac{d\theta }{2\pi }\sum_{j=0}^{2\sigma }\left( 
\begin{array}{c}
2\sigma  \\ 
j
\end{array}
\right) \exp \left\{ 2i\theta \left( j-\sigma \right) \right\} ,  \label{a6}
\end{equation}
and hence, after summation over $j$ and some rearrangements, 
\begin{equation}
\Lambda _{2\sigma }^{\left( N\right) }=\frac 2\pi \int_0^{{\cal D}_N}\frac{%
dt\,t^{2\sigma }}{\left( {\cal D}_N^2-t^2\right) ^{1/2}}.  \label{a7}
\end{equation}
Here ${\cal D}_N=2c_N.$

The integral 
\begin{equation}
\Gamma _{2\sigma +1}^{\left( N\right) }=\int d\alpha \left( t\right)
P_N^{\left( \alpha \right) }\left( t\right) P_{N-1}^{\left( \alpha \right)
}\left( t\right) t^{2\sigma +1}  \label{a8}
\end{equation}
with integer $\sigma \geq 0$ is computable in the same manner, with an
answer given by 
\begin{equation}
\Gamma _{2\sigma +1}^{\left( N\right) }=\frac 2{\pi {\cal D}_N}\int_0^{{\cal %
D}_N}\frac{dt\,t^{2\sigma +2}}{\left( {\cal D}_N^2-t^2\right) ^{1/2}}.
\label{a9}
\end{equation}
Notice that due to the asymptotic property $c_{N+q}\approx c_N$ mentioned
above, $\Lambda _{2\sigma }^{\left( N+q\right) }\approx \Lambda _{2\sigma
}^{\left( N\right) }$ and $\Gamma _{2\sigma +1}^{\left( N+q\right) }\approx $
$\Gamma _{2\sigma +1}^{\left( N\right) }$ for $q\sim {\cal O}\left(
N^0\right) $.

Finally, we demonstrate the usefulness of the formulas (\ref{a7}) and (\ref{a9}) by
finding the explicit expressions for the functions $A_{\rm{reg}}^{\left(
N\right) }\left( \lambda \right) $ and $B_{\rm{reg}}^{\left( N\right)
}\left( \lambda \right) $, defined by Eqs. (\ref{eq.a02}) and (\ref{eq.b02}%
). Substitution of Eq. (\ref{a7}) into Eq. (\ref{eq.a02}) followed by summation
over $\sigma $ yields 
\begin{eqnarray}
A_{\rm{reg}}^{\left( N\right) }\left( \lambda \right)  &=&\frac{2{\cal D}_N%
}\pi {\cal P}\int_0^{{\cal D}_N}\frac{dt\,}{\left( {\cal D}_N^2-t^2\right)
^{1/2}}\sum_{k=1}^pd_k\frac{t^{2k}-\lambda ^{2k}\,}{t^2-\lambda ^2} 
\nonumber \\
&=&\frac{2{\cal D}_N}\pi {\cal P}\int_0^{{\cal D}_N}\frac{dt\,}{\left( {\cal %
D}_N^2-t^2\right) ^{1/2}}\frac{tdv/dt}{t^2-\lambda ^2}.  \label{a10}
\end{eqnarray}
Analogously, we obtain 
\begin{eqnarray}
B_{\rm{reg}}^{\left( N\right) }\left( \lambda \right)  &=&\frac 2\pi {\cal %
P}\int_0^{{\cal D}_N}\frac{tdt\,}{\left( {\cal D}_N^2-t^2\right) ^{1/2}}%
\frac{\lambda dv/dt-tdv/d\lambda }{t^2-\lambda ^2}  \nonumber \\
&=&\frac{2\lambda }\pi {\cal P}\int_0^{{\cal D}_N}\frac{dt\,}{\left( {\cal D}%
_N^2-t^2\right) ^{1/2}}\frac{tdv/dt}{t^2-\lambda ^2}-\frac{dv}{d\lambda }.
\label{a11}
\end{eqnarray}
One can convince himself that Eqs. (\ref{a10}) and (\ref{a11}) obey the sum rule Eq. (%
\ref{eq.29}) for the confinement potential $V_\alpha $, Eq. (\ref{v}), and
for the functions $A_N^{\left( \alpha \right) }$ and $B_N^{\left( \alpha \right)
}$, given by Eqs. (\ref{eq.A}) and (\ref{eq.B}). Notice that this is the recurrence
equation (\ref{eq.16}) which enabled us to obtain closed analytic
expressions (\ref{a10}) and (\ref{a11}) relating the functions $A_{\rm{reg}}^{\left(
N\right) }$ and $B_{\rm{reg}}^{\left( N\right) }$ to the regular part $v$
of the confinement potential.

\section{Integral Representation of $\Lambda _{2\sigma }^{\left( N\right) }$
and $\Gamma _{2\sigma +1}^{\left( N\right) }$: Two--Band Phase}

Consider the integral 
\begin{equation}
\Lambda _{2\sigma }^{\left( N\right) }=\int d\alpha \left( t\right)
P_N^2\left( t\right) t^{2\sigma }  \label{b1}
\end{equation}
with integer $\sigma \geq 0$. It follows from Eq. (\ref{tbq.19}) that in the
two--band phase the following asymptotic identities exist, 
\begin{eqnarray}
\lambda ^{2m}P_N\left( \lambda \right)  &=&\left(
c_N^2+c_{N-1}^2\right) ^m\sum_{k=0}^m\left( 
\begin{array}{c}
m \\ 
k
\end{array}
\right) \left( \frac{c_Nc_{N-1}}{c_N^2+c_{N-1}^2}\right) ^k  \nonumber
\label{b2} \\
&&\times \sum_{j=0}^k\left( 
\begin{array}{c}
k \\ 
j
\end{array}
\right) P_{N+4j-2k}\left( \lambda \right) ,  \label{b2}
\end{eqnarray}
and 
\begin{eqnarray}
\lambda ^{2m+1}P_N\left( \lambda \right)  &=&\left(
c_N^2+c_{N-1}^2\right) ^m\sum_{k=0}^m\left( 
\begin{array}{c}
m \\ 
k
\end{array}
\right) \left( \frac{c_Nc_{N-1}}{c_N^2+c_{N-1}^2}\right)
^k\sum_{j=0}^k\left( 
\begin{array}{c}
k \\ 
j
\end{array}
\right)   \nonumber \\
&&\times \left[ c_{N-1}P_{N+4j-2k+1}\left( \lambda \right)
+c_NP_{N+4j-2k-1}\left( \lambda \right) \right]   \label{b3}
\end{eqnarray}
with integer $m\geq 0$. Both Eqs. (\ref{b2}) and (\ref{b3}) can be proven by the
mathematical induction. Making use of Eq. (\ref{b2}) we rewrite $\Lambda _{2\sigma
}^{\left( N\right) }$ in the form 
\begin{eqnarray}
\Lambda _{2\sigma }^{\left( N\right) } &=&\left( c_N^2+c_{N-1}^2\right)
^\sigma \sum_{k=0}^\sigma \left( 
\begin{array}{c}
\sigma  \\ 
k
\end{array}
\right) \left( \frac{c_Nc_{N-1}}{c_N^2+c_{N-1}^2}\right) ^k  \nonumber
\label{b4} \\
&&\times \sum_{j=0}^k\left( 
\begin{array}{c}
k \\ 
j
\end{array}
\right) \int d\alpha \left( t\right) P_N\left( t\right) P_{N+4j-2k}\left(
t\right) .  \label{b4}
\end{eqnarray}
Orthogonality of $P_n$ allows us to integrate over the measure $d\alpha $,
thus simplifying Eq. (\ref{b4}) to 
\begin{equation}
\Lambda _{2\sigma }^{\left( N\right) }=\left( c_N^2+c_{N-1}^2\right) ^\sigma
\sum_{k=0}^\sigma \left( 
\begin{array}{c}
\sigma  \\ 
k
\end{array}
\right) \left( \frac{c_Nc_{N-1}}{c_N^2+c_{N-1}^2}\right)
^k\sum_{j=0}^k\left( 
\begin{array}{c}
k \\ 
j
\end{array}
\right) \delta _{2j}^k.  \label{b5}
\end{equation}
Substituting the integral representation (\ref{a5}) for the Kronecker symbol, and
performing the double summation over indices $j$ and $k$, we obtain 
\begin{equation}
\Lambda _{2\sigma }^{\left( N\right) }=\int_0^{2\pi }\frac{d\theta }{2\pi }%
\left( c_N^2+c_{N-1}^2+2c_Nc_{N-1}\cos \theta \right) ^\sigma \rm{.}
\label{b6}
\end{equation}
Introducing a new integration variable $t^2=c_N^2+c_{N-1}^2+2c_Nc_{N-1}\cos
\theta $, we derive the final formula 
\begin{equation}
\Lambda _{2\sigma }^{\left( N\right) }=\frac 2\pi \int_{{\cal D}_N^{-}}^{%
{\cal D}_N^{+}}\frac{t^{2\sigma +1}dt}{\left[ \left( {\cal D}_N^{+}\right)
^2-t^2\right] ^{1/2}\left[ t^2-\left( {\cal D}_N^{-}\right) ^2\right] ^{1/2}}
\label{b7}
\end{equation}
with 
\begin{equation}
{\cal D}_N^{\pm }=\left| c_N\pm c_{N-1}\right| .  \label{b8}
\end{equation}

The integral 
\begin{equation}
\Gamma _{2\sigma +1}^{\left( N\right) }=\int d\alpha \left( t\right)
P_N\left( t\right) P_{N-1}\left( t\right) t^{2\sigma +1}  \label{b9}
\end{equation}
with integer $\sigma \geq 0$ is evaluated in the same way. Making use of
expansion Eq. (\ref{b3}), we rewrite Eq. (\ref{b9}) in the form that allows us to
perform the integration over the measure $d\alpha $, 
\begin{eqnarray}
\Gamma _{2\sigma +1}^{\left( N\right) } &=&\frac 12\left(
c_N^2+c_{N-1}^2\right) ^\sigma \int d\alpha \left( t\right) P_{N-1}\left(
t\right) \sum_{k=0}^\sigma \left( 
\begin{array}{c}
\sigma  \\ 
k
\end{array}
\right) \left( \frac{c_Nc_{N-1}}{c_N^2+c_{N-1}^2}\right) ^k  \nonumber \\
&&\ \times \sum_{j=0}^k\left( 
\begin{array}{c}
k \\ 
j
\end{array}
\right) \left[ c_{N-1}P_{N+4j-2k+1}\left( t\right) +c_NP_{N+4j-2k-1}\left(
t\right) \right] .  \label{b10}
\end{eqnarray}
After integration, we get 
\begin{eqnarray}
\Gamma _{2\sigma +1}^{\left( N\right) } &=&\frac 12\left(
c_N^2+c_{N-1}^2\right) ^\sigma \sum_{k=0}^\sigma \left( 
\begin{array}{c}
\sigma  \\ 
k
\end{array}
\right) \left( \frac{c_Nc_{N-1}}{c_N^2+c_{N-1}^2}\right) ^k  \nonumber \\
&&\times \sum_{j=0}^k\left( 
\begin{array}{c}
k \\ 
j
\end{array}
\right) \left[ c_{N-1}\delta _{2j+1}^k+c_N\delta _{2j}^k\right] .
\label{b11}
\end{eqnarray}
The double summation in Eq. (\ref{b11}) can be performed by using the integral
representation for the Kronecker symbol given by Eq. (\ref{a5}), 
\begin{equation}
\Gamma _{2\sigma +1}^{\left( N\right) }=\frac 12\int_0^{2\pi }\frac{d\theta 
}{2\pi }\left( c_N^2+c_{N-1}^2+2c_Nc_{N-1}\cos \theta \right) ^\sigma \left[
c_N+c_{N-1}\cos \theta \right] .  \label{b12}
\end{equation}
Introducing a new integration variable $t^2=c_N^2+c_{N-1}^2+2c_Nc_{N-1}\cos
\theta $, we get 
\begin{equation}
\Gamma _{2\sigma +1}^{\left( N\right) }=\frac 1{\pi c_N}\int_{{\cal D}%
_N^{-}}^{{\cal D}_N^{+}}\frac{t^{2\sigma +1}dt}{\left[ \left( {\cal D}%
_N^{+}\right) ^2-t^2\right] ^{1/2}\left[ t^2-\left( {\cal D}_N^{-}\right)
^2\right] ^{1/2}}\left[ t^2+c_N^2-c_{N-1}^2\right] .  \label{b13}
\end{equation}
Notice that because $P_{-1}\left( \lambda \right) =0$, it follows from
Eq. (\ref{eq.16}) that $c_0=0$, and as a consequence, an even branch $c_{2N}$
always lies lower than an odd branch $c_{2N\pm 1}$, so that $c_{2N}<c_{2N\pm
1}$. Then, we may conclude from Eq. (\ref{b8}) that 
\begin{equation}
c_N=\frac{{\cal D}_N^{+}-\left( -1\right) ^N{\cal D}_N^{-}}2,  \label{b14}
\end{equation}
and, as a consequence, 
\begin{equation}
\Gamma _{2\sigma +1}^{\left( N\right) }=\frac 1{\pi c_N}\int_{{\cal D}%
_N^{-}}^{{\cal D}_N^{+}}\frac{t^{2\sigma +1}dt}{\left[ \left( {\cal D}%
_N^{+}\right) ^2-t^2\right] ^{1/2}\left[ t^2-\left( {\cal D}_N^{-}\right)
^2\right] ^{1/2}}\left[ t^2-\left( -1\right) ^N{\cal D}_N^{-}{\cal D}%
_N^{+}\right] .  \label{b15}
\end{equation}

\section{Soft Edges in the Two--Band Phase}

To find the equations determining the end points ${\cal D}_N^{\pm }$ where
the Dyson spectral density goes to zero, we start with the following formula
from the theory of orthogonal polynomials \cite{N-1986} 
\begin{equation}
n=2c_n\int d\alpha \left( t\right) \frac{dV}{dt}P_n\left( t\right)
P_{n-1}\left( t\right) ,  \label{c1}
\end{equation}
also known as a ``string equation''. Let us use expansion Eq. (\ref{b3}) to
evaluate the integral entering Eq. (\ref{c1}) in the limit $n=N\gg 1$. It is easy
to see that 
\begin{equation}
N=2c_N\sum_{k=1}^pd_k\int d\alpha \left( t\right) P_N\left( t\right)
P_{N-1}\left( t\right) t^{2k-1}=2c_N\sum_{k=1}^pd_k\Gamma _{2k-1}^{\left(
N\right) },  \label{c2}
\end{equation}
where $\Gamma _{2k-1}^{\left( N\right) }$ is given by Eq. (\ref{b13}). Then, we
immediately obtain the relationship 
\begin{equation}
N=\frac 2\pi \int_{{\cal D}_N^{-}}^{{\cal D}_N^{+}}\frac{dt}{\left[ \left( 
{\cal D}_N^{+}\right) ^2-t^2\right] ^{1/2}\left[ t^2-\left( {\cal D}%
_N^{-}\right) ^2\right] ^{1/2}}\frac{dV}{dt}\left[
t^2+c_N^2-c_{N-1}^2\right] .  \label{c3}
\end{equation}
This result, rewritten for $n=N-1$, yields in the large--$N$ limit, 
\begin{equation}
N=\frac 2\pi \int_{{\cal D}_N^{-}}^{{\cal D}_N^{+}}\frac{dt}{\left[ \left( 
{\cal D}_N^{+}\right) ^2-t^2\right] ^{1/2}\left[ t^2-\left( {\cal D}%
_N^{-}\right) ^2\right] ^{1/2}}\frac{dV}{dt}\left[
t^2+c_{N-1}^2-c_N^2\right] .  \label{c4}
\end{equation}
Equations (\ref{c3}) and (\ref{c4}) bring us two integral equations whose solutions
determine the end points ${\cal D}_N^{\pm }$, 
\begin{equation}
\int_{{\cal D}_N^{-}}^{{\cal D}_N^{+}}\frac{t^2dt}{\left[ \left( {\cal D}%
_N^{+}\right) ^2-t^2\right] ^{1/2}\left[ t^2-\left( {\cal D}_N^{-}\right)
^2\right] ^{1/2}}\frac{dV}{dt}=\frac{\pi N}2,  \label{c5}
\end{equation}
and 
\begin{equation}
\int_{{\cal D}_N^{-}}^{{\cal D}_N^{+}}\frac{dt}{\left[ \left( {\cal D}%
_N^{+}\right) ^2-t^2\right] ^{1/2}\left[ t^2-\left( {\cal D}_N^{-}\right)
^2\right] ^{1/2}}\frac{dV}{dt}=0.  \label{c6}
\end{equation}
As ${\cal D}_N^{-}\rightarrow 0$, Eq. (\ref{c5}) coincides with the integral
equation (\ref{mrs}) for a single--band phase. In the same limit, Eq. (\ref{c6})
becomes equivalent to the assertion $\nu _D\left( 0\right) =0$, with $\nu _D$
being the spectral density in a single--band phase. This corresponds to the
point of merging of two eigenvalue cuts.


\begin{thebibliography}{}
\bibitem{M-1991} Mehta, M. L. (1991) {\it Random Matrices},  Academic Press,
Boston.

\bibitem{GMGW-1998} Guhr, T., M\"{u}ller--Groeling, A., and Weidenm\"{u}ller, H. A. (1998) Random matrix theories in quantum physics: Common
concepts, {\it Phys. Rep.} {\bf 299}, 189 -- 428.

\bibitem{W-1951}  Wigner, E. P. (1951) On the statistical distribution of the
widths and spacings of nuclear resonance levels, {\it Proc. Cambridge Philos.
Soc.} {\bf 47}, 790 -- 798.

\bibitem{D-1962}  Dyson, F. J. (1962) Statistical theory of energy levels of
complex systems I, II, III, {\it J. Math. Phys.} {\bf 3}, 140 -- 156; {\bf 3},
157 -- 165; {\bf 3}, 166 -- 175.

\bibitem{D-1962a} Dyson, F. J. (1962) The threefold way. Algebraic structure
of symmetry groups and ensembles in quantum mechanics,  {\it J. Math. Phys.} {\bf 3%
}, 1199 -- 1215.

\bibitem{BFFMPW-1981} Brody, T. A., Flores, J., French, J. B., Mello, P. A., Pandey, A., and Wong, S. S. M. (1981) Random--matrix physics: Spectrum and
strength fluctuations, {\it Rev. Mod. Phys.} {\bf 53}, 385 -- 479.

\bibitem{FGZJ-1995}  Di Francesco, P., Ginsparg, P., and Zinn--Justin, J. (1995)
$2D$ gravity and random matrices, {\it Phys. Rep.} {\bf 254}, 1 -- 133.

\bibitem{VZ-1993} Verbaarschot, J. J. M. and Zahed, I. (1993) Spectral density
of the QCD Dirac operator near zero virtuality, {\it Phys. Rev. Lett.} {\bf 70},
3852 -- 3855; Verbaarschot, J. J. M. (1994) The spectrum of the QCD Dirac
operator and chiral random matrix theory: the threefold way, {\it Phys. Rev.
Lett.} {\bf 72}, 2531 -- 2533; Verbaarschot, J. J. M. and Zahed, I. (1994) Random
matrix theory and QCD, {\it Phys. Rev. Lett.} {\bf 73}, 2288 -- 2291.

\bibitem{G-1990} Gutzwiller, M. C. (1990) {\it Chaos in Classical and Quantum
Mechanics}, Springer--Verlag, New York.

\bibitem{ALW-1991} Altshuler, B. L., Lee, P. A., and Webb, R. A. (1991) {\it Mesoscopic Phenomena in Solids}, North--Holland, Amsterdam.

\bibitem{B-1997}  Beenakker, C. W. J. (1997) Random matrix theory of quantum
transport, {\it Rev. Mod. Phys.} {\bf 69}, 731 -- 808.

\bibitem{GE-1965} Gor'kov, L. P. and Eliashberg, G. M. (1965) Minute metallic
particles in an electromagnetic field, {\it Zh. Eksp. Teor. Fiz.} {\bf 48}, 1407 -- 1418 [{\it Sov. Phys. JETP} {\bf 21}, 940 -- 947].

\bibitem{P-1965} Porter, C. E. (1965) {\it Statistical theories of spectra:
Fluctuations}, Academic Press, New York.

\bibitem{E-1983}  Efetov, K. B. (1983) Supersymmetry and theory of disordered
metals, {\it Adv. Phys.} {\bf 32}, 53 -- 127.

\bibitem{BBMSVW-1998} Berbenni--Bitsch, M. E., Meyer, S., Sch\"{a}fer, A., Verbaarschot, J. J. M., and Wettig, T. (1998) Microscopic universality in the
spectrum of the lattice Dirac operator, {\it Phys. Rev. Lett.} {\bf 80}, 1146 --1149.

\bibitem{QCD-1998} Verbaarschot, J. J. M. (1998) Universal fluctuations in
Dirac spectra, in {\it New Developments in Quantum Field Theory}, Plenum
Press, New York.

\bibitem{ADMN-1997} Akemann, G., Damgaard, P. H., Magnea, U., and Nishigaki, S. (1997) Universality of random matrices in the microscopic limit and
the Dirac operator spectrum, {\it Nucl. Phys. B} {\bf 487}, 721 -- 738.

\bibitem{D-1972} Dyson, F. J. (1972) A class of matrix ensembles, {\it J. Math.
Phys.} {\bf 13}, 90 -- 97.

\bibitem{AJM-1990} Ambj\o rn, J., Jurkiewicz, J., and Makeenko, Yu. M. (1990) Multiloop correlators for two--dimensional quantum gravity, {\it Phys. Lett. B} 
{\bf 251}, 517 -- 524.

\bibitem{AA-1996} Akemann, G. and Ambj\o rn, J. (1996) New universal spectral
correlators, {\it J. Phys. A} {\bf 29}, L555 -- L560.

\bibitem{I-1997}  Itoi, C. (1997) Universal wide correlators in non--Gaussian
orthogonal, unitary and symplectic random matrix ensembles, {\it Nucl. Phys. B}
{\bf 493}, 651 -- 659.

\bibitem{B-1993} Beenakker, C. W. J. (1993) Universality in the
random--matrix theory of quantum transport, {\it Phys. Rev. Lett.} {\bf 70}, 1155 -- 1158.

\bibitem{B-1994} Beenakker, C. W. J. (1994) Universality of Br\'{e}zin and
Zee's spectral correlator, {\it Nucl. Phys. B} {\bf 422}, 515 -- 520.

\bibitem{BZ-1994} Br\'{e}zin, E. and Zee, A. (1994) Correlation functions in
disordered systems, {\it Phys. Rev. E} {\bf 49}, 2588 -- 2596.

\bibitem{HW-1995} Hackenbroich, G. and Weidenm\"{u}ller, H. A. (1995) Universality of random--matrix results for non--Gaussian ensembles, {\it Phys.
Rev. Lett.} {\bf 74}, 4118 -- 4121.

\bibitem{VWZ-1985} Verbaarschot, J. J. M., Weidenm\"{u}ller, H. A., and Zirnbauer, M. R. (1985) Grassmann integration in stochastic quantum physics: The
case of compound--nucleus scattering, {\it Phys. Rep.} {\bf 129}, 367 -- 438.

\bibitem{GM-1960} Mehta, M. L. (1960) On the statistical properties of the
level--spacings in nuclear spectra, {\it Nucl. Phys.} {\bf 18}, 395 -- 419; Mehta, M. L. and
Gaudin, M. (1960) On the density of eigenvalues of a random matrix, {\it Nucl. Phys.} {\bf 18}, 420 -- 427.

\bibitem{FK-1964} Fox, D. and Kahn, P. B. (1964) Higher order spacing
distributions for a class of unitary ensembles, {\it Phys. Rev.} {\bf 134}, B1151 -- B1155.

\bibitem{L-1964} Leff, H. S. (1964) Class of ensembles in the statistical
theory of energy--level spectra, {\it J. Math. Phys.} {\bf 5}, 763 -- 768.

\bibitem{B-1965} Bronk, B. V. (1965) Exponential ensemble for random matrices, {\it J. Math. Phys.} {\bf 6}, 228 -- 237.

\bibitem{NW-1991} Nagao, T. and Wadati, M. (1991) Correlation functions of
random matrix ensembles related to classical orthogonal polynomials I, {\it J. Phys. Soc. Jpn.} {\bf 60}, 3298 -- 3322; Nagao, T. and Wadati, M. (1992) Correlation functions of
random matrix ensembles related to classical orthogonal polynomials II, III, {\it J. Phys. Soc. Jpn.} {\bf 61}, 78 -- 88; {\bf 61}, 1910 -- 1918.

\bibitem{BTW-1992} Basor, E. L., Tracy, C. A., and Widom, H. (1992) Asymptotics
of level spacing distribution functions for random matrices, {\it Phys. Rev.
Lett.} {\bf 69}, 5 -- 8.

\bibitem{BN-1991} Br\'{e}zin, E. and Neuberger, H. (1991) Multicritical points
of unoriented random surfaces, {\it Nucl. Phys. B} {\bf 350}, 513 -- 553.

\bibitem{TW-1998} Tracy, C. A. and Widom, H. (1998) Correlations functions,
cluster functions and spacing distributions for random matrices, {\it Los Alamos
preprint archive}, solv--int/9804004.

\bibitem{MM-1991} Mahoux, G. and Mehta, M. L. (1991) A method of integration
over matrix variables: IV, {\it J. Phys. (France)} {\bf 1}, 1093 -- 1108.

\bibitem{BZ-1993} Br\'{e}zin, E. and Zee, A. (1993) Universality of the
correlations between eigenvalues of large random matrices, {\it Nucl. Phys. B}
{\bf 402}, 613 -- 627.

\bibitem{FKY-1996} Freilikher, V., Kanzieper, E., and Yurkevich, I. (1996) Unitary random--matrix ensemble with governable level confinement, {\it Phys.
Rev. E} {\bf 53}, 2200 -- 2209.

\bibitem{FKY-1996a} Freilikher, V., Kanzieper, E., and Yurkevich, I. (1996) Theory of random matrices with strong level confinement: Orthogonal
polynomial approach, {\it Phys. Rev. E} {\bf 54}, 210 -- 219.

\bibitem{ST-1963} Shohat, J. A. and Tamarkin, J. D. (1963) {\it The problem of
moments}, American Mathematical Society, Providence.

\bibitem{S-1921} Szeg\"{o}, G. (1921) \"{U}ber die Randwerte analytischer
Funktionen, {\it Math. Annalen} {\bf 84}, 232 -- 244.

\bibitem{SN-1993}  Slevin, K. and Nagao, T. (1993) New random matrix theory of
scattering in mesoscopic systems, {\it Phys. Rev. Lett.} {\bf 70}, 635 -- 638.

\bibitem{AZ-1997} Altland, A. and Zirnbauer, M. R. (1997) Nonstandard symmetry
classes in mesoscopic normal--superconducting hybrid structures, {\it Phys. Rev.
B} {\bf 55}, 1142 -- 1161.

\bibitem{NS-1993} Nagao, T. and Slevin, K. (1993) Nonuniversal correlations
for random matrix ensembles, {\it J. Math. Phys.} {\bf 34}, 2075 -- 2085;
Laguerre ensembles of random matrices: Nonuniversal correlation functions,
{\it J. Math. Phys.} {\bf 34}, 2317 -- 2330.

\bibitem{F-1993} Forrester, P. J. (1993) The spectrum of random matrix
ensembles, {\it Nucl. Phys. B} {\bf 402}, 709 -- 728.

\bibitem{NF-1995} Nagao, T. and Forrester, P. J. (1995) Asymptotic
correlations at the spectrum edge of random matrices, {\it Nucl. Phys. B} {\bf 435%
}, 401 -- 420.

\bibitem{TW-1994} Tracy, C. A. and Widom, H. (1994) Level--spacing
distributions and the Bessel kernel, {\it Commun. Math. Phys.} {\bf 161}, 289 -- 309.

\bibitem{N-1996} Nishigaki, S. (1996) Proof of universality of the Bessel
kernel for chiral matrix models in the microscopic limit, {\it Phys. Lett. B}
{\bf 387}, 139 -- 144.

\bibitem{KF-1998} Kanzieper, E. and Freilikher, V. (1998) Random--matrix
models with the logarithmic--singular level confinement: Method of
fictitious fermions, {\it Philos. Mag. B} {\bf 77}, 1161 -- 1171.

\bibitem{W-1962}  Wigner, E. P. (1962) Distribution laws for the roots of a
random Hermitean matrix, reprinted in: Porter, C. E. (1965) {\it %
Statistical theories of spectra: Fluctuations}, pp. 446 -- 461 , Academic Press, New
York.

\bibitem{BB-1991} Bowick, M. J. and Br\'{e}zin, E. (1991) Universal scaling of
the tail of the density of eigenvalues in random matrix models, {\it Phys. Lett.
B} {\bf 268}, 21 -- 28.

\bibitem{KF-1997} Kanzieper, E. and Freilikher, V. (1997) Universality in
invariant random--matrix models: Existence near the soft edge, {\it Phys. Rev. E}
{\bf 55}, 3712 -- 3715.

\bibitem{TW-1994a} Tracy, C. A. and Widom, H. (1994) Level--spacing
distributions and the Airy kernel, {\it Commun. Math. Phys.} {\bf 159}, 151 -- 174.

\bibitem{KF-1997a} Kanzieper, E. and Freilikher, V. (1997) Novel universal
correlations in invariant random--matrix models, {\it Phys. Rev. Lett.} {\bf 78},
3806 -- 3809.

\bibitem{DN-1998} Damgaard, P. H. and Nishigaki, S. M. (1998) Universal
spectral correlators and massive Dirac operators, {\it Nucl. Phys. B} {\bf %
518}, 495 -- 512; Universal massive spectral correlators and QCD$_3$, {\it Phys. Rev. D} {\bf 57}, 5299 -- 5302.

\bibitem{A-1996} Akemann, G. (1996) Higher genus correlators for the
Hermitian matrix model with multiple cuts, {\it Nucl. Phys. B} {\bf 482}, 403 -- 430.

\bibitem{KF-1998a} Kanzieper, E. and Freilikher, V. (1998) Two--band random
matrices, {\it Phys. Rev. E} {\bf 57}, 6604 -- 6611.

\bibitem{S-1939} Shohat, J. (1930) {\it C. R. Hebd. Seances Acad. Sci.} {\bf 191},
989; Shohat, J. (1939) A differential equation for orthogonal polynomials, {\it Duke
Math. J.} {\bf 5}, 401 -- 417.

\bibitem{SV-1998} Sener, M. K. and Verbaarschot, J. J. M. (1998) Universality
in chiral random matrix theory at $\beta =1$ and $\beta =4$, {\it Los
Alamos preprint archive}, hep--th/9801042.

\bibitem{W-1998} Widom, H. (1998) On the relation between orthogonal,
symplectic and unitary matrix ensembles, {\it Los Alamos preprint archive},
solv--int/9804005.

\bibitem{S-1967} Szeg\"{o}, G. (1967) {\it Orthogonal Polynomials}, American
Mathematical Society, Providence.

\bibitem{BC-1986} Bonan, S. S. and Clark, D. S. (1986) Estimates of the
orthogonal polynomials with weight $\exp (-x^m)$, $m$ an even positive
integer, {\it J. Appr. Theory} {\bf 46}, 408 -- 410.

\bibitem{JATs} Sheen, R. C. (1987) Plancherel--Rotach type asymptotics for
orthogonal polynomials associated with $\exp \left( -x^6/6\right)$, {\it J.
Appr. Theory} {\bf 30}, 232 -- 293; Bonan, S. S. and Clark, D. S. (1990) Estimates of the Hermite and the Freud polynomials, {\it J. Appr. Theory } {\bf 63}%
, 210 -- 224; Bauldry, W. C. (1990) Estimates of asymmetric Freud polynomials, 
{\it J. Appr. Theory } {\bf 63}, 225 -- 237; Mhaskar, H. N. (1990) Bounds for
certain Freud type orthogonal polynomials, {\it J. Appr. Theory} {\bf 63}, 238 -- 254.

\bibitem{N-1986} Nevai, P. (1986) G\'{e}za Freud, orthogonal polynomials and
Christoffel functions: A case study,  {\it J. Appr. Theory} {\bf 48}, 3 -- 167.

\bibitem{L-1993}  Lubinsky, D. S. (1993) An update on orthogonal polynomials
and weighted approximation on the real line, {\it Acta Appl. Math.} {\bf 33}, 121 -- 164.

\bibitem{AK-1998} Akemann, G. and Kanzieper, E. (1998) unpublished.

\bibitem{ADMN-1998} Akemann, G., Damgaard, P. H., Magnea, U., and
Nishigaki, S. (1998) Multicritical microscopic spectral correlators of Hermitian
and complex matrices, {\it Nucl. Phys. B} {\bf 519}, 682 -- 714.

\bibitem{GM-1990} Gross, D. J. and Migdal, A. A. (1990) Nonperturbative
two--dimensional quantum gravity, {\it Phys. Rev. Lett.} {\bf 64}, 127 -- 130.

\bibitem{D-1973} Dingle, R. B. (1973) {\it Asymptotic Expansions: Their Derivation
and Interpretation}, Academic Press, New York.

\bibitem{M-1988} Molinari, L. (1988) Phase structure of matrix models through
orthogonal polynomials, {\it J. Phys. A} {\bf 21}, 1 -- 6.

\bibitem{DSS-1990} Douglas, M., Seiberg, N., and Shenker, S. (1990) Flow and
instability in quantum gravity, {\it Phys. Lett. B} {\bf 244}, 381 -- 386.

\bibitem{SS-1991} Sasaki, M. and Suzuki, H. (1991) Matrix realization of
random surfaces, {\it Phys. Rev. D} {\bf 43}, 4015 -- 4028.

\bibitem{JNZ-1996}  Jurkiewicz, J., Nowak, M. A., and Zahed, I. (1996) Dirac
spectrum in QCD and quark masses, {\it Nucl. Phys. B} {\bf 478}, 605 -- 626.

\bibitem{JV-1996} Jackson, A. and Verbaarschot, J. (1996) Random matrix model
for chiral symmetry breaking, {\it Phys. Rev. D} {\bf 53}, 7223 -- 7230.

\bibitem{CKPR-1995} Cugliandolo, L. F., Kurchan, J., Parisi, G., and
Ritort, F. (1995) Matrix models as solvable glass models, {\it Phys. Rev. Lett.} {\bf %
74}, 1012 -- 1015.

\bibitem{MHK-1995} Morita, Y., Hatsugai, Y., and Kohmoto, M. (1995) Universal
correlations in random matrices and one--dimensional particles with
long--range interactions in a confinement potential, {\it Phys. Rev. B} {\bf 52},
4716 -- 4719. See also Ref. \cite{HINS-1997}.

\bibitem{S-1982} Shimamune, Y. (1982) On the phase structure of large $N$ 
matrix models and gauge models, {\it Phys. Lett. B} {\bf 108}, 407 -- 410.

\bibitem{CMM-1990} Cicuta, G. M., Molinari, L., and Montaldi, E. (1990) Multicritical points in matrix models, {\it J. Phys. A} {\bf 23}, L421 -- L425.

\bibitem{J-1991} Jurkiewicz, J. (1991) Chaotic behaviour in one--matrix models, {\it Phys. Lett. B} {\bf 261}, 260 -- 268.

\bibitem{DDJT-1990}  Demeterfi, K., Deo, N., Jain, S., and Tan, C.--I. (1990)
Multiband structure and critical behavior of matrix models, {\it Phys. Rev. D} 
{\bf 42}, 4105 -- 4122.

\bibitem{HINS-1997} Higuchi, S., Itoi, C., Nishigaki, S. M., and Sakai, N. (1997)
Renormalization group approach to multiple--arc random matrix models,
{\it Phys. Lett. B} {\bf 398}, 123 -- 129.

\bibitem{D-1997}  Deo, N. (1997) Orthogonal polynomials and exact correlation
functions for two cut random matrix models, {\it Nucl. Phys. B} {\bf 504}, 609 --
620.

\bibitem{AS-1986} Altshuler, B. L. and Shklovskii, B. I. (1986) Repulsion of
energy levels and conductivity of small metal samples, {\it Zh. Eksp. Teor. Fiz.}
{\bf 91}, 220 -- 234 [{\it Sov. Phys. JETP} {\bf 64}, 127 -- 135].

\bibitem{JPB-1993} Jalabert, R. A., Pichard, J.--L., and Beenakker, C. W. J. (1993)
Long--range energy level interaction in small metallic particles, {\it Europhys. Lett.} {\bf 24}, 1 -- 6.

\bibitem{BD-1998} Br\'{e}zin, E. and Deo, N. (1998) Smoothed correlators for
symmetric double--well matrix models: Some puzzles and resolutions, {\it Los
Alamos preprint archive}, cond--mat/9805096.

\bibitem{BZJ-1992}  Br\'{e}zin, E. and Zinn--Justin, J. (1992) Renormalization
group approach to matrix models, {\it Phys. Lett. B} {\bf 288}, 54 -- 58.

\bibitem{MCIN-1993} Muttalib, K. A., Chen, Y., Ismail, M. E. H., and
Nicopoulos, V. N. (1993) New family of unitary random matrices, {\it Phys. Rev. Lett. }
{\bf 71}, 471 -- 475.

\bibitem{BBP-1997} Bogomolny, E., Bohigas, O., and Pato, M. P. (1997)
Distribution of eigenvalues of certain matrix ensembles, {\it Phys. Rev. E} {\bf %
55}, 6707 -- 6718.
\end{thebibliography}
\end{document}